\documentclass[twocolumn,  % Toggle 2-column style
			   trackchanges]{aastex701}
			   
\usepackage{graphicx}
\usepackage{comment}
\usepackage{multirow}
\usepackage{amsmath}
\usepackage{amssymb}
\usepackage{mathtools}
\usepackage[utf8]{inputenc}

\usepackage[dvipsnames]{xcolor}

\newcommand{\mchirp}{\mathcal{M}_\mathrm{chirp}}
\newcommand{\mdet}{\mathcal{M}_\mathrm{det}}
\newcommand{\R}{R_\odot}
\newcommand{\M}{\,M_\odot}
\newcommand{\mmax}{M_\mathrm{max}}
\newcommand{\mmin}{M_\mathrm{min}}
\newcommand{\rmax}{R_\mathrm{max}}

\newcommand{\erf}{\mathrm{erf}}
\newcommand{\AN}{\mathrm{AN}}
\newcommand{\SN}{\mathrm{SN}}
\newcommand{\vbar}{\,|\,}

\renewcommand{\(}{\left(}
\renewcommand{\)}{\right)}

\begin{document}

\title{Inference of Neutron Star Mass Distributions and the Dense Matter Equation of State from Multi-messenger Observations}

\author[orcid=0000-0003-0256-083X,
		gname=Mahmudul Hasan,
		sname=Anik]{Mahmudul Hasan Anik}
\affiliation{The University of Tennessee at Knoxville, 1408 Circle Drive, Knoxville TN 37996, USA}
\email{anik@vols.utk.edu}

\author[orcid=0000-0003-2478-4017,
		gname=Andrew,
		sname=Steiner]{Andrew W. Steiner}
\affiliation{Eureka Scientific, Inc., 2452 Delmer Street, Oakland, CA 94602, USA}
\email{awsteiner0@protonmail.com}

\author[orcid=0000-0001-5832-8517,
		gname=Richard,
		sname=O'Shaughnessy]{Richard O'Shaughnessy}
\affiliation{Rochester Institute of Technology, 1 Lomb Memorial Drive, Rochester, NY 14623, USA}
\email{rossma@rit.edu}

\begin{abstract}
We construct a combined model to incorporate neutron star (NS) mass
measurements with electromagnetic mass-radius constraints and
gravitational-wave observations using Bayesian inference. We use
different mass distributions for three populations depending on the
companion stars: double neutron stars, NS - white dwarfs, and low-mass
X-ray binaries (LMXB). To observe the effects of different
parametrizations, we use two equation of state (EoS) models: a
piecewise polytrope and a fixed sound-speed model at high densities, in
combination with a low-density EoS. Our results show that the mass
distributions of these NS populations are distinct and sensitive to
the EoS prior choices. In addition, we show for the first time that 
using a uniform prior on the observable NS maximum mass, rather than 
a nuisance parameter in the unknown high-density EoS, shifts the 
posterior maximum mass to larger values. For polytropic EoSs, the 
maximum mass posterior changes from $\mmax=2.09_{-0.07}^{+0.18}\M$ to 
$2.15_{-0.10}^{+0.19}\M$ at 90\% confidence level. This change in 
prior also impacts the shape of the mass distribution for NSs in LMXB, 
shifting the posterior for the population mean from 
$\mu_{\rm lmxb}=1.51_{-0.13}^{+0.13}\M$ to $1.62_{-0.12}^{+0.15}\M$ at 
68\% confidence level.
%, changing theskewness from negative to positive values.

\end{abstract}



\section{Introduction} \label{sec:intro}

Neutron stars (NSs) provide insight to the strong interactions of dense matter in extreme conditions 
governed by the nuclear equation of state (EoS). The EoS at low densities is well understood within the 
context of nuclear physics, for example, from the chiral effective field theory ($\chi$EFT) at densities 
below $2n_0$ \citep{keller-23, tews-25}, where $n_0=0.16$ fm$^{-3}$ is the nuclear saturation density. 
Moreover, nuclear experiments such as neutron-skin thickness measurements of $^{108}$Pb and $^{48}$Ca by 
PREX-II \citep{adhikari-21} and CREX \citep{adhikari-22}, respectively, provide important information on the 
symmetry energy and its slope which also constrain the low-density EoS. However, the EoS is dominated by 
large uncertainties at higher densities and heavily relies on astrophysical observations. In this era of 
multi-messenger astronomy, an increasingly growing number of NS observational data from various sources have 
become available since the last decade. This includes mass-radius constraints from electromagnetic (EM) 
observations of globular clusters \citep{steiner-18}, type-I X-ray bursters \citep{nattila-16}, NICER X-ray 
observations \citep{riley-19, miller-21}, mass measurements from pulsar radio timing \citep{alsing-18}, and 
the LIGO-Virgo observations of gravitational waves (GWs) \citep{ligo-17, ligo-19}. Particularly, while 
precisely measured masses from more observations of massive pulsars impose strict lower bounds on the NS 
maximum mass ($\mmax$) and thus the EoS, perturbative quantum chromodynamics (pQCD) at extremely high 
densities ($\ge 40n_0$) provides reliable theoretical constraints on the EoS from above 
\citep{komoltsev-22}. Therefore, combining models to incorporate nuclear theory and experiments with NS 
observational data from multi-messenger astronomy are crucial to constraining the dense matter EoS and 
revealing the underlying NS mass distribution.

Efforts in modeling the NS mass distributions using Bayesian inference began long before the era of GW and 
NICER observations \citep{finn-94, thorsett-99, schwab-10, valentim-11}. In a novel approach, \cite{ozel-12} 
categorized NS mass measurements based on the available information for NS binaries and modeled the mass 
distribution using a single Gaussian for each type. They inferred the peaks at $1.28 \pm 0.24 \M$ for 
eclipsing high-mass X-ray binaries and slow pulsars (near birth masses), $1.33 \pm 0.05 \M$ for double 
neutron stars (DNS), and $1.48 \pm 0.2 \M$ for recycled NSs. On the other hand, \cite{kiziltan-13} used a 
skewed normal distribution for NS binaries depending on the companion types and found a tight symmetric 
distribution for DNS with the peaks at $1.33 \M$ for DNS and a slightly skewed distribution (high-mass tail) 
with the peak at $1.55 \M$ for NS-WD binaries. The authors also observed a mass cutoff at $\sim 2.1 \M$ for 
NS-WD which they suggested should be a lower bound for the maximum NS mass ($\mmax$). 

In a later work focused on millisecond pulsars (MSPs), \cite{antoniadis-16} found that the MSP mass 
distribution is strongly asymmetric which they argued is best accounted for by a truncated bimodal 
distribution. They inferred the low-mass peak at $1.393 \pm 0.064 \M$ and the high-mass peak at $1.807 \pm 
0.177 \M$, and $\mmax \ge 2.018 \M$ at 98\% confidence level (CL), closely matching the prediction by 
\cite{kiziltan-13}. In their analysis, \cite{alsing-18} presented another approach by using a $n-$component 
Gaussian mixture with a cutoff at $\mmax$ for the combined population. Using NS mass data across DNS, NS-WD, 
X-ray binaries, and a model selection method, the authors found evidence for bimodality ($n=2$) and a sharp 
cutoff with $2.0 < \mmax/\M < 2.2$ at 68\% CL, which is nearly insensitive to the model choice and the most 
massive stars in the data. This study led future works to include massive pulsars to investigate their 
influences on the NS mass distributions, specifically the cutoff at $\mmax$. For instance, \cite{farr-20} 
included the pulsar J0740+6620 with mass $\sim 2.14 \M$ \citep{cromartie-20} and observed no significant 
change in the $\mmax$ posterior but a weaker cutoff than reported in \cite{alsing-18}, attributed to the 
choice of $\mmax$ prior. Another extension is \cite{shao-20} who increased the sample size by adding data 
from recent studies and found $\mmax = 2.26_{-0.05}^{+0.12} \M$ at 68\% CL, which suggested their influence 
on the $\mmax$ posterior. Other works also using the bimodal Gaussian function to model the NS mass 
distribution either used synthetic data \citep{farrow-19, chatziioannou-20, golomb-22}, or limited their 
analyses to GW observations (DNS) and NS-BH \citep{landry-21, li-21}. Note that the $\mmax$ posteriors in 
these works are inferred only from the mass distributions themselves and thus uninformed by any EoS models.

Results from the literature discussed above clearly shows that the inferred NS mass distribution exhibits 
bimodality only when all galactic NSs are collectively viewed as a single population. It is also evident 
that the $\mmax$ posterior solely informed by the mass distributions (with no EoS inputs) can change with 
the inclusion of more precisely measured massive stars. In \cite{wysocki-20}, the authors demonstrated that 
the NS mass distribution and the nuclear EoS must be inferred simultaneously to avoid bias arising from 
independent analyses. \cite{golomb-22}, also highlighted the importance of jointly inferring mass 
distribution and EoS rather than treating them individually. Recently \cite{fan-24} adopted a hybrid 
approach by inferring $\mmax$ from the NS population which was then used to reconstruct the EoS models for 
joint inference. They also used constraints from $\chi$EFT and pQCD at low and high densities, respectively, 
and significantly large samples of NS masses across including black widow and redback MSPs. The authors 
found that $\mmax = 2.25_{-0.07}^{+0.08} \M$ at 68\% CL. Using the same data sets additionally combined with 
the PREX-II and CREX measurements, \cite{biswas-25a} obtained $\mmax = 2.22_{-0.19}^{+0.21} \M$ at 90\% CL. 
Most recently, \cite{golomb-25} modeled the EM observations of galactic NSs using the same bimodal Gaussian 
and GW observations with a power law. The authors assigned an astrophysical maximum mass ($M_\mathrm{pop}$) 
to each of the two populations for truncation assuming that they may differ from the $\mmax$ supported by 
the EoS (where $M_\mathrm{pop} < \mmax$), and claimed that doing so would allow them to investigate whether 
the NS maximum mass in different populations is limited by the EoS or the astrophysical processes. However, 
they found no evidence of the two maximum masses being different and found $\mmax = 2.28_{-0.21}^{+0.41} \M$ 
at 90\% CL.

In this work, we construct a combined model to incorporate GW and EM observations with NS mass distributions 
using Bayesian inference. Our data sets include the GW observations (GW170817 and GW190425), mass-radius 
constraints from quiescent low-mass X-ray binaries (qLMXB) in globular clusters, photospheric radius 
expansion (PRE) X-ray bursters, NICER observations of J0740+6620 and the isolated pulsar J0030+0451, and 
mass measurements from radio timing, X-ray, and optical observations of 58 NSs in binaries where the 
individual NS masses are known. Each NS (except J0030+0451) in our data belongs to either of the three NS 
binaries: DNS, NS-WD, and LMXB. Assuming NS populations have different mass distributions depending on the 
companion stars, we assign the mass distribution models to each of them. Next, we examine the effects of 
different EoS parametrizations and prior choices by using two hybrid models - a low-density EoS combined 
with a piecewise polytrope and a fixed sound-speed model at higher densities. As a result, our joint 
inference obtains distinct mass distributions for each of the three NS populations along with posteriors for 
$\mmax$. Finally, we demonstrate, for the first time, how prior assumptions on the EoS-informed $\mmax$ 
influence the nature of observable matter at the highest densities.

In the following Section \ref{sec:tf}, we describe our data sets and present our models for the NS mass 
distribution, the EoS, and the GW observations. Then in Section \ref{sec:ba}, we discuss the parameters and 
prior choices in our Bayesian inference. Next, in Section \ref{sec:rd}, we present our results and compare 
with recent works. Finally in Section \ref{sec:con}, we conclude by highlighting the implications of our 
results in the context of previous studies in this area.

\section{Theoretical Framework \label{sec:tf}} 
Throughout this work, neutron stars are presumed non-rotating, non-accreting, isotropic, and spherically 
symmetric objects. Furthermore, our EoS models are simple and uninformative of the microscopic nuclear 
interactions and possible phase transitions. In this section, we begin by briefly discussing the data sets, 
and then explain the NS mass distribution, the equations of state, and the GW models.

\subsection{NS Populations and Data} \label{sub:data}

\begin{deluxetable*}{l c r}
	\tablecaption{Stars in DNS with measured masses and probability distributions (${\cal P}$). 
		\label{tab:dns}}
	\tablehead{\colhead{Star} & \colhead{Mass [$M_\odot$]/Data} & \colhead{Reference}}
	\startdata
	J0453+1559      & $1.559_{-0.004}^{+0.004}$     & \citet{martinez-15} \\
	J0453+1559 c.   & $1.174_{-0.004}^{+0.004}$     & \citet{martinez-15} \\
	J1906+0746      & $1.291_{-0.011}^{+0.011}$     & \citet{vanleeuwen-15} \\
	J1906+0746 c.   & $1.322_{-0.011}^{+0.011}$     & \citet{vanleeuwen-15} \\
	B1534+12        & $1.3332_{-0.001}^{+0.001}$    & \citet{fonseca-14} \\
	B1534+12 c.     & $1.3452_{-0.001}^{+0.001}$    & \citet{fonseca-14} \\
	B1913+16        & $1.4398_{-0.0002}^{+0.0002}$  & \citet{weisberg-10} \\
	B1913+16 c.     & $1.3886_{-0.0002}^{+0.0002}$  & \citet{weisberg-10} \\
	B2127+11C       & $1.358_{-0.01}^{+0.01}$       & \citet{jacoby-06} \\
	B2127+11C c.    & $1.354_{-0.01}^{+0.01}$       & \citet{jacoby-06} \\
	J0737-3039A     & $1.3381_{-0.0007}^{+0.0007}$  & \citet{kramer-06} \\
	J0737-3039B     & $1.2489_{-0.0007}^{+0.0007}$  & \citet{kramer-06} \\
	J1756-2251      & $1.312_{-0.017}^{+0.017}$     & \citet{ferdman-14} \\
	J1756-2251 c.   & $1.258_{-0.017}^{+0.017}$     & \citet{ferdman-14} \\
	J1807-2500B     & $1.3655_{-0.0021}^{+0.0021}$  & \citet{lynch-12} \\
	J1807-2500B c.  & $1.2064_{-0.002}^{+0.002}$    & \citet{lynch-12} \\
	J1518+4904      & $1.56_{-0.44}^{+0.13}$        & \citet{thorsett-99} \\
	J1518+4904 c.   & $1.05_{-0.11}^{+0.45}$        & \citet{thorsett-99} \\
	J1811-1736      & $1.56_{-0.45}^{+0.24}$        & \citet{stairs-06, corongiu-07} \\
	J1811-1736 c.   & $1.12_{-0.13}^{+0.47}$        & \citet{stairs-06, corongiu-07} \\
	J1829+2456      & $1.20_{-0.46}^{+0.12}$        & \citet{champion-05} \\
	J1829+2456 c.   & $1.40_{-0.12}^{+0.46}$        & \citet{champion-05} \\
	GW170817 $m_1$  & ${\cal P}~(\mathcal{M}_\mathrm{chirp}, \tilde{\Lambda}, q)$ & \citet{ligo-17} \\
	GW170817 $m_2$  & ${\cal P}~(\mathcal{M}_\mathrm{chirp}, \tilde{\Lambda}, q)$ & \citet{ligo-17} \\
	GW190425 $m_1'$ & ${\cal P}~(m_1')$             & \citet{ligo-19} \\
	GW190425 $m_2'$ & ${\cal P}~(m_2')$             & \citet{ligo-19}
	\enddata
	\tablecomments{The symbol ``c." following a star's name indicates the companion.}
\end{deluxetable*}

Our primary assumption here is that NS masses observed in each type of binary, based on the companion star, 
follow a different underlying mass distribution. We also assume that the mass distributions do not evolve 
with time and are independent of the different evolution paths of the individual stars.

This work investigates three populations depending on the companions: NS-NS or double neutron star (DNS), 
NS--white dwarf (NS-WD), and low-mass X-ray binary (LMXB). The NS data sets include gravitational wave (GW) 
observations, electromagnetic (EM) mass--radius constraints, and mass measurements from radio timing, X-ray, 
and optical observations. Our analysis excludes NS binaries where the individual NS masses are not directly 
measured but include constraints on the mass function, the total mass, and/or the mass ratio.

The stars in DNS, except the GWs, have precisely measured masses. The GW170817 data is a 3-dimensional 
probability density of the chirp mass, tidal deformability, and mass ratio. The GW190425 data, however, 
contains simple mass probability densities with no information on tidal deformation or redshift. Note that 
we do not distinguish between galactic DNS and merging NS binaries, nor do we separate them based on their 
available data or detection methods. Consequently, the GW stars are presumed to follow the same DNS mass 
distribution as other stars in this population. On the other hand, NS-WD contains mass measurements of 
several massive pulsars ($\ge 1.8 \M$) with generally wider error bars than DNS. It also includes the NICER 
observation of J0740+6620 with mass--radius constraints. Next, LMXB includes the EM mass-radius data from 
quiescent low-mass X-ray binaries (qLMXB) in globular clusters and photospheric radius expansion X-ray 
bursters (PREs), in addition to a few other stars that have mass measurements with large error bars. 
Finally, we also include the isolated pulsar PSR J0030+0451 which is treated separately from the mass 
distribution models.

The full lists of NS binaries in our data are given in Tables \ref{tab:dns}-\ref{tab:lmxb} grouped by 
populations. The mass measurements are reported within 68\% central limits, along with their source 
references. There are 26 stars in DNS, 32 in NS-WD, and 16 in LMXB.

\begin{deluxetable*}{l c r}[htbp]
	\tablecaption{Stars in NS--WD and measured masses with 68\% central limits.\label{tab:nswd}}
	\tablehead{
		\colhead{Star} & \colhead{Mass [$M_\odot$]/Data} & \colhead{Reference}
	}
	\startdata
	J2045+3633    & $1.33_{-0.3}^{+0.3}$           & \citet{berezina-17} \\
	J2053+4650    & $1.4_{-0.21}^{+0.21}$          & \citet{berezina-17} \\
	J1713+0747    & $1.35_{-0.07}^{+0.07}$         & \citet{arzoumanian-18} \\
	B1855+09      & $1.37_{-0.13}^{+0.13}$         & \citet{arzoumanian-18} \\
	J0751+1807    & $1.72_{-0.07}^{+0.07}$         & \citet{desvignes-16} \\
	J1141-6545    & $1.27_{-0.01}^{+0.01}$         & \citet{bhat-08} \\
	J1738+0333    & $1.47_{-0.07}^{+0.07}$         & \citet{antoniadis-12} \\
	J1614-2230    & $1.908_{-0.016}^{+0.016}$      & \citet{arzoumanian-18} \\
	J0348+0432    & $2.01_{-0.04}^{+0.04}$         & \citet{antoniadis-13} \\
	J2222-0137    & $1.76_{-0.06}^{+0.06}$         & \citet{cognard-17} \\
	J2234+0611    & $1.393_{-0.013}^{+0.013}$      & \citet{stovall-19} \\
	J1949+3106    & $1.47_{-0.43}^{+0.43}$         & \citet{deneva-12} \\
	J1012+5307    & $1.83_{-0.11}^{+0.11}$         & \citet{antoniadis-16} \\
	J0437-4715    & $1.44_{-0.07}^{+0.07}$         & \citet{reardon-16} \\
	J1909-3744    & $1.48_{-0.03}^{+0.03}$         & \citet{arzoumanian-18} \\
	J1802-2124    & $1.24_{-0.11}^{+0.11}$         & \citet{ferdman-10} \\
	J1911-5958A   & $1.34_{-0.08}^{+0.08}$         & \citet{bassa-06} \\
	J2043+1711    & $1.38_{-0.13}^{+0.13}$         & \citet{arzoumanian-18} \\
	J0337+1715    & $1.4378_{-0.0013}^{+0.0013}$   & \citet{ransom-14} \\
	J1946+3417    & $1.828_{-0.022}^{+0.022}$      & \citet{barr-17} \\
	J1918-0642    & $1.29_{-0.1}^{+0.1}$           & \citet{arzoumanian-18} \\
	J1600-3053    & $2.3_{-0.7}^{+0.7}$            & \citet{arzoumanian-18} \\
	J0621+1002    & $1.7_{-0.17}^{+0.10}$          & \citet{nice-08} \\
	B2303+46      & $1.38_{-0.1}^{+0.06}$          & \citet{thorsett-99} \\
	J0024-7204H   & $1.48_{-0.06}^{+0.03}$         & \citet{kiziltan-13} \\
	J0514-4002A   & $1.49_{-0.27}^{+0.04}$         & \citet{kiziltan-13} \\
	B1516+02B     & $2.1_{-0.19}^{+0.19}$          & \citet{kiziltan-13} \\
	J1748-2446I   & $1.91_{-0.1}^{+0.02}$          & \citet{kiziltan-13} \\
	J1748-2446J   & $1.79_{-0.1}^{+0.02}$          & \citet{kiziltan-13} \\
	B1802-07      & $1.26_{-0.17}^{+0.08}$         & \citet{thorsett-99} \\
	B1911-5958A   & $1.4_{-0.10}^{+0.16}$          & \citet{bassa-06} \\
	J0740+6620    & ${\cal P}(M, R)$               & \citet{riley-21, miller-21}
	\enddata
\end{deluxetable*}

\begin{deluxetable*}{l c r}[htbp]
	\tablecaption{Stars in LMXB and measured masses and probability distributions (${\cal P}$), and the 
		isolated pulsar PSR J0030+0451.\label{tab:lmxb}}
	\tablehead{
		\colhead{Star} & \colhead{Mass [$M_\odot$]/Data} & \colhead{Reference}
	}
	\startdata
	Cyg X-2          & $1.71_{-0.21}^{+0.21}$   & \citet{casares-10} \\
	XTE J2123-058    & $1.53_{-0.42}^{+0.42}$   & \citet{gelino-02} \\
	4U 1822-371      & $1.96_{-0.36}^{+0.36}$   & \citet{munozdarias-05} \\
	Her X-1          & $1.073_{-0.36}^{+0.36}$  & \citet{rawls-11} \\
	2S 0921-630      & $1.44_{-0.10}^{+0.10}$   & \citet{steeghs-07} \\
	47 Tuc (X7)      & ${\cal P}(M, R)$         & \citet{steiner-18} \\
	$\omega$ Cen     & ${\cal P}(M, R)$         & \citet{steiner-18} \\
	NGC 6304         & ${\cal P}(M, R)$         & \citet{steiner-18} \\
	NGC 6397         & ${\cal P}(M, R)$         & \citet{steiner-18} \\
	M13              & ${\cal P}(M, R)$         & \citet{steiner-18} \\
	M28              & ${\cal P}(M, R)$         & \citet{steiner-18} \\
	M30              & ${\cal P}(M, R)$         & \citet{steiner-18} \\
	SAX J1810.8-2609 & ${\cal P}(M, R)$         & \citet{nattila-16} \\
	4U 1702-429      & ${\cal P}(M, R)$         & \citet{nattila-17} \\
	4U 1724-307      & ${\cal P}(M, R)$         & \citet{nattila-16} \\
	\hline
	J0030+0451       & ${\cal P}(M, R)$         & \citet{riley-19, miller-19}
	\enddata
	%\tablecomments{}
\end{deluxetable*}

\subsection{Mass Distribution Model} \label{sub:mdist}
We assume that the NS mass distributions differ by populations, but remain unchanged for each population. In 
other words, the individual neutron stars may undergo different evolution paths, but their overall mass 
distribution in a given population does not evolve.

For the $i$-th star, the measured mass $m_i$, which may differ from the NS mass $M_i$ by $w_i$, is defined as
\begin{equation} \label{eq:mod}
	m_i \equiv M_i + w_i,
\end{equation}
where $i = 1, \dots , n$, and assume that $M_i$ is a sample drawn from the skewed normal distribution given 
by
\begin{equation} \label{eq:sn}
	\text{SN}(M \vbar \mu, \sigma, \alpha) = \frac{2}{\sigma} ~ \phi\left(\frac{M-\mu}{\sigma}\right) ~ \Phi\left[\frac{(M-\mu)\alpha}{\sigma}\right], 
\end{equation}
where $\phi$, $\Phi$ are the standard normal and the cumulative distribution functions, and $\mu$, $\sigma$, 
$\alpha$ are location, scale, and skewness parameters, respectively. Note that Equation (\ref{eq:sn}) 
becomes a normal distribution for $\alpha=0$, right-skewed for $\alpha>0$, and left-skewed for $\alpha<0$.

To model the asymmetric error bars in the $i$-th NS mass measurement ${m_i}_{-l_i}^{+u_i}$, where $-l_i, 
+u_i$ are the 68\% central limits and $l_i \neq u_i$, we assume that $w_i$ is drawn from an asymmetric 
normal distribution given by
\begin{eqnarray} \label{eq:an}
	\AN(w \vbar c, d) &=& \frac{2}{d\(c+1/c\)} \nonumber \\ 
	&\times& \left[\phi\left(\frac{w}{cd}\right) \Theta(w)
	+ \phi\left(\frac{cw}{d}\right) \Theta(-w)\right],
\end{eqnarray}
where $\Theta$ is the Heaviside step function and $c>0$, $d>0$ are constants which for the $i$-th star can 
be calculated as $c_i=\sqrt{u_i/l_i}$, and then solving for $d_i$:

\begin{equation}
	\int_{-l_i}^{u_i} \AN(w \vbar c_i, d_i) \, dw = 0.68.
\end{equation}

Note that Equation (\ref{eq:an}) becomes a normal distribution for the symmetric case $c=1$ (i.e., when 
$l_i=u_i$), right-skewed for $c>1$, and left-skewed for $c<1$. The details of these calculations are given 
in Appendix \ref{app:a}.

\subsection{EoS Models} \label{sub:eos}
We use a numerical EoS for the crust at densities $0 \leq n_b < 0.04$ fm$^3$, and an EoS adopted from 
\cite{gandolfi-12} at $0.04$ fm$^3 \leq n_b < 2n_0$, where $n_0=0.16$ fm$^3$ is the nuclear saturation 
density, which is referred to as the low-density EoS. Above $n_b \ge 2n_0$, we use two EoS models - the 
linear EoS from \cite{steiner-13} and a piecewise polytrope.

The low-density EoS is constructed for pure neutron matter based on 2- and 3-nucleon interactions using 
quantum Monte Carlo techniques. The pressure is given by
\begin{equation} \label{eq:ld}
	P(n_b) = n_b\left[a \alpha \(\frac{n_b}{n_0}\)^\alpha + b \beta \(\frac{n_b}{n_0}\)^\beta\right],
\end{equation}
where the parameters $a, \alpha, b, \beta$ are related to the symmetry energy $S$ and its derivative $L$ by
\begin{equation} \label{eq:sl}
	S = a + b + 16.0, \quad L = 3(a\alpha + b\beta).
\end{equation}
The symmetry energy and its derivative are correlated and further constrained \citep{steiner-15} by,
\begin{equation}
	(9.17S - 266.0 \ \mathrm{MeV}) < L < (14.3S - 379.0 \ \mathrm{MeV}).
\end{equation} 

The linear EoS, denoted as NL, contains three line segments on the $P-\epsilon$ plane where each segment has 
a fixed speed of sound $c_s$ and a fixed energy density $\epsilon_0$ at $P=0$. It is also piecewise 
continuous and given by
\begin{equation} \label{eq:nl}
	P(\epsilon) = c_s^2(\epsilon-\epsilon_0),
\end{equation}
where $c_s^2$ is the relative speed of sound squared defined as $c_s^2 = dP/d\epsilon$. 

The polytropic EoS, denoted as NP, is piecewise continuous and consists of three polytropes connected at the 
transition densities. The pressure as a function of the energy density is given by
\begin{equation} \label{eq:np}
	P(\epsilon) = K\epsilon^\gamma,
\end{equation}
where $K$ is a proportionality constant and $\gamma$ is the adiabatic index.

Note that both the polytropic and linear EoS models are simple $P-\epsilon$ relationships and completely 
uninformative of nuclear interactions and possible phase transitions.

\subsection{Gravitational Wave Models \label{sub:gw}}
For the GW170817 data, the likelihood, taken from \cite{mamun-21}, is a function of the detector-frame chirp 
mass $\mdet$, mass ratio $q \equiv m_2/m_1 < 1$, and tidal deformability $\tilde{\Lambda}$. This likelihood 
is interpolated from the marginal likelihood provided by RIFT \citep{lange-18} and integrated over the two 
dimensionless NS spins $\chi_{1,z}$, $\chi_{2,z}$ relative to the orbital angular momentum direction, where 
$z$ is the redshift. Given the model parameters $\mdet$, $q$, $z$, we compute the chirp mass
\begin{equation}
	\mchirp = \frac{\mdet}{1+z},
\end{equation}
and then the individual NS masses $m_1$, $m_2$ as follows:
\begin{subequations}
\begin{align}
	m_1 &= \frac{\mchirp(1+q)^{1/5}}{q^{3/5}} \label{eq:m1} \\
	m_2 &= \mchirp \ q^{2/5} (1+q)^{1/5} \label{eq:m2}.
\end{align}
\end{subequations}
Next, we extract their radii from the EoS, compute the moments of inertia $I_{1,2}$, and then the tidal 
deformabilities $\Lambda_{1,2}$ using the fitting method from \cite{steiner-16}:
\begin{equation}
	\ln \Lambda_k = \sum_{i=0}^4 b_i (\ln I_k)^i,
\end{equation} 
where $b_i$ are the fitting coefficients and $k=1,2$. This method avoids direct calculations of tidal deformabilities from the EoS, which is computationally expensive.

The dimensionless combined tidal deformability is then given by
\begin{equation}
	\tilde{\Lambda} = \frac{16}{13}~\frac{(m_1+12m_2)m_1^4\Lambda_1 + (m_2+12m_1)m_2^4\Lambda_2}{(m_1+m_2)^5}.
\end{equation}

For the GW190425 data, we use its precisely measured chirp mass as a constant to compute the likelihood 
$\mathcal{P}(m_1^\prime)$ given the model parameter $m_1^\prime$, where the mass ratio $q^\prime$ and 
$m_2^\prime$ are given by (\ref{eq:m1}-\ref{eq:m2}).

\section{Bayesian Analysis \label{sec:ba}}
We employ Bayesian inference to determine credible intervals for our model parameters using the 
observational data. Bayes's theorem states that the posterior distribution of parameter $\theta$ given data 
$d$ is
\begin{equation}
	\mathcal{P}(\theta|d) = C \mathcal{L}(d|\theta) \, \mathcal{P}(\theta),
\end{equation}
where $\mathcal{L}(d|\theta)$ is the likelihood of the observation $d$ given the parameter $\theta$, 
$\mathcal{P}(\theta)$ is the prior distribution of $\theta$, and $C$ is the normalization constant.

The models and likelihoods have already been described in Section \ref{sec:tf}. Next, we discuss the 
parameters and their prior distributions.

\subsection{Parameters and Prior Choices \label{sub:ppc}} 
The skewed normal distribution given by Equations (\ref{eq:sn}) for each population requires the shape 
parameters mean $\mu$, width $\sigma$, and skewness $\alpha$. Thus, there are three sets of the parameters 
$(\mu_i, \sigma_i, \alpha_i)$ where $i=1, 2, 3$ for the populations DNS, NS-WD, and LMXB, respectively, a 
total of 9 shape parameters. 

Additionally to obtain posterior distributions for the individual stars, we assign a mass parameter 
$M_{i,j}$ for the $j$-th star in the $i$-th population that has a mass measurement. In total, there are 
71 mass parameters. As explained in Subsection \ref{sub:mdist}, GW170817 is modeled with its detector-frame chirp mass $\mdet$, mass ratio $q$, and redshift $z$. As for GW190425, the only parameter is mass $m_1$. Hence, there are 4 parameters for the two GW observations. 

Finally, our parametrized hybrid EoS has total 9 parameters. The low-density EoS given by Equation 
(\ref{eq:ld}) requires 4 parameters: coefficient $a$, exponent $\alpha$, symmetry energy $S$, and its slope 
$L$. The polytropic EoS has 5 parameters: the exponents $\gamma_1$, $\gamma_2$, $\gamma_3$ - one for each of 
the three polytropes, and the transition densities $\epsilon_1$ and $\epsilon_2$ between them. The linear EoS 
also has 5 parameters with the same transition densities, but the exponents $\gamma_k$ are replaced by the 
relative speed of sound $c_{s,k}^2$, where $k=1,2,3$ for three line segments. 

\begin{deluxetable}{l c c c}[t]
	\tablecaption{List of all parameters with their symbols, units, and uniform prior choices.}
	\label{tab:prior}
	\tablehead{
		\colhead{Parameter} & \colhead{Unit} & \colhead{Low} & \colhead{High}
	}
	\startdata
	Mean, $\mu_i$                      & $M_\odot$   & $0.5$     & $2.5$     \\
	Width, $\log_{10}\sigma_i$         &             & $-6.0$    & $0.0$     \\
	Skewness, $\alpha_i$               &             & $-1.0$    & $1.0$     \\
	Mass, $M_{i,j}$                    & $M_\odot$   & $1.0$     & $2.5$     \\
	%Mass fraction, $\mathrm{mf}_{i,j}$ &             & $0.0$     & $1.0$     \\
	Coefficient, $a$                   & MeV         & $12.5$    & $13.5$    \\
	Exponent, $\alpha$                 &             & $0.47$    & $0.53$    \\
	Symmetry energy, $S$               & MeV         & $29.5$    & $36.1$    \\
	S.E. Slope, $L$                    & MeV         & $30.0$    & $70.0$    \\
	Exponent, $\gamma_k$               &             & $10^{-6}$ & $10.0$    \\
	Sound speed, $c_{s,k}^2$           &             & $0.0$     & $1.0$     \\
	Transition density, $\epsilon_{1,2}$ & fm$^{-4}$   & $0.75$    & $8.0$     \\
	Chirp mass, $\mdet$                & $M_\odot$   & $1.1917$  & $1.1979$  \\
	Mass ratio, $q$                    &             & $0.0$     & $1.0$     \\
	Redshift, $z$                      &             & $0.0$     & $1.0$     \\
	\enddata
	\tablecomments{Index $i$ is over 3 populations, and $j=1, \dots, N_i$, where $N_i$ is the number of NSs in the $i$-th population.}
\end{deluxetable} 

During sampling the EoS parameters, we check the causality and stability conditions: $0<c_s(\epsilon)<1$. 
In addition, we ensure that the transition densities are ordered and do not exceed the central density of 
the maximum-mass star, i.e., $\epsilon_1<\epsilon_2<\epsilon_\mathrm{max}$. If any of these conditions are violated, the EoS is discarded.

Thus, there are 92 parameters which are listed in Table
\ref{tab:prior} along their prior choices. We choose uniform
distributions for all parameters, including the widths ($\sigma_i$)
which are flat in log-space. Our prior choice is that the two (polytrope-like vs. linear) EoS models carry 
equal weights.

\subsection{Combined Model} \label{sub:imd}
For clarity, the number of stars with measured masses in each population is $n_i$, where $i=1$ for DNS, 
$i=2$ for NS-WD, and $i=3$ for LMXB. The number of stars with EM mass--radius data is $n_\mathrm{em}$ and 
number of GW stars is $n_\mathrm{gw}$. Thus, the total number of neutron stars in our data is 
$N=n_1+n_2+n_3+n_\mathrm{gw}+n_\mathrm{em}$, where $n_1+n_\mathrm{gw}$ stars are in DNS, $n_2+1$ in NS-WD, 
and $n_3+n_\mathrm{em}-1$ in LMXB.

Next, as a reminder, $m_{i,j}$ are the measured NS masses, $c_{i,j}$ and $d_{i,j}$ are constants computed 
from the mass data (see Subsection \ref{sub:mdist}). Finally, The 9 EoS parameters are collectively denoted 
by $\{p\}$, including both the low- and high-density EoS models. For explanations of all other symbols 
representing the parameters, see Table \ref{tab:prior}. The combined likelihood function is

\begin{widetext}
\begin{eqnarray} \label{eq:wgt}
	\mathcal{L} && ~\left(\{\mu_i\}, \{\sigma_i\}, \{\alpha_i\}, \{M_{i,j}\}, \mdet, q, z, m_1^\prime, \{p\}\right) \nonumber \\ 
	&& =\prod_{i=1}^3 \prod_{j=1}^{n_i} \AN\(m_{i,j}-M_{i,j}, c_{i,j}, d_{i,j}\) ~ \SN\(\mu_i, \sigma_i, \alpha_i, M_{i,j}\) ~ \Theta\(M_{\mathrm{max}}-M_{i,j}\) ~ \Theta\(M_{i,j}-M_{\mathrm{min}}\) \nonumber \\
	&& \times \prod_{i=2}^{3} \prod_{j=1}^{n_\mathrm{em}} \mathcal{L}_\mathrm{em}\left[R(M_{i,j}, \{p\}), M_{i,j}\right] ~ \SN\(\mu_i, \sigma_i, \alpha_i, M_{i,j}\) ~ \Theta\(M_{\mathrm{max}}-M_{i,j}\) ~ \Theta\(M_{i,j}-M_{\mathrm{min}}\) \nonumber \\ 
	&& \times \prod_{j=1}^{n_\mathrm{gw}} \SN\(\mu_1, \sigma_1, \alpha_1, M_{1,j}\) ~ \Theta\(M_{\mathrm{max}}-M_{1,j}\) ~ \Theta\(M_{1,j}-M_{\mathrm{min}}\) \nonumber \\
	&& \times ~ \mathcal{L}_\mathrm{gw17} \left[\mchirp(\mdet, z), ~q, ~\tilde{\Lambda}(\mdet, q, \{p\})\right] ~ \mathcal{L}_\mathrm{gw19} (m_1^\prime),
\end{eqnarray}
\end{widetext}
where $\Theta$ is the Heaviside function, $\mmin \equiv 1.0 \M$ is the NS minimum mass, and $\mmax$ is the 
maximum mass supported by the EoS and thus a function of the EoS parameters $\{p\}$.

\subsection{{Changing parameters}}
We want to see how the results change when we make the maximum mass a parameter with a flat prior 
distribution. We do this by making a change of variable. Before the transformation, the likelihood is of the 
form ${\cal L}(\ldots, p_1,\ldots,p_{N_k-1},p_{N_k})$ given above in Equation (\ref{eq:wgt}) and the prior, which we assume to be a 
product of independent factors, is

\begin{equation}
	\prod_i {\cal P}(\mu_i) {\cal P}(\sigma_i) {\cal P}(\alpha_i)
	\prod_{i,j} {\cal P}(M_{i,j})  \prod_k {\cal P}(p_k),
\end{equation}

where $i$ is 1, 2, or 3, $j$ runs over all the neutron stars in each
class, and $k$ indexes the EoS parameters. Our goal is to modify the
prior distribution to the new form

\begin{equation}
	\prod_i {\cal P}(\mu_i) {\cal P}(\sigma_i) {\cal P}(\alpha_i)
	\prod_{i,j} {\cal P}(M_{i,j})  \prod_{k=1}^{N_k-1} {\cal P}(p_k)
	{\cal P}(M_{\mathrm{max}}),
\end{equation}
which we can do with the identity
\begin{equation}
	{\cal P}(M_{\mathrm{max}}) = \left|\frac{\partial p_{N_k}
		(\{p_1,\ldots,p_{N_k-1}\},M_{\mathrm{max}}) }
	{\partial M_{\mathrm{max}}} \right|~{\cal P}(p_{N_k}).
\end{equation}
Thus, the new likelihood is
\begin{eqnarray}
	&{\cal L}&^\prime~(\ldots,p_1,\ldots,p_{N_k-1},M_{\mathrm{max}}) \nonumber \\ 
	&=& {\cal L}~(\ldots,p_1,\ldots,p_{N_k-1},p_{N_k})~\left|\frac{\partial p_{N_k}}{\partial
		M_{\mathrm{max}}}\right|_{\{p_1,\ldots,p_{N_k-1}\}}.
\end{eqnarray}
In the implementation, we sample $p_{N_k}$ directly and include the Jacobian factor in the MCMC target density at each step. For the linear model, $p_{N_k}$ is the normalized speed of sound squared in the third segment, $c_{s,3}^2$, while for the polytropic model, $p_{N_k}$ is the exponent of the third polytrope, $\gamma_3$. Holding $p_1,\ldots,p_{N_k-1}$ fixed, $p_{N_k}$ uniquely determines $M_{\mathrm{max}}$, so either variable may be used to parameterize the same EoS space. Therefore, sampling $p_{N_k}$ and computing $M_{\mathrm{max}}$, or equivalently sampling $M_{\mathrm{max}}$ and recovering the corresponding $p_{N_k}$, gives the same reparameterized posterior when the Jacobian factor is included. 

The Jacobian is evaluated numerically by perturbing $p_{N_k}$ while keeping all other EoS parameters fixed,
\begin{equation}
	p_{N_k}' = p_{N_k} + \epsilon p_{N_k}, \quad 0<\epsilon\ll 1.
\end{equation}
We set $\epsilon=0.01$, which provides a good balance between accuracy and numerical stability. The EoS is reconstructed with $p_{N_k}'$, and the TOV equations are solved again to obtain the corresponding perturbed maximum mass $M_{\mathrm{max}}'$. The Jacobian is then approximated as
\begin{equation}
	\left|\frac{\partial p_{N_k}}{\partial M_{\mathrm{max}}}\right|_{\{p_1,\ldots,p_{N_k-1}\}} \approx \left|\frac{p_{N_k}'-p_{N_k}}{M_{\mathrm{max}}'-M_{\mathrm{max}}}\right|_{\{p_1,\ldots,p_{N_k-1}\}}.
\end{equation}
On rare occasions, numerical noise can give $M_{\mathrm{max}}'\le M_{\mathrm{max}}$, in which case the EoS is rejected. This keeps the finite-difference estimate of the Jacobian finite and well defined. Since $p_{N_k}$ is sampled directly, no root-finding or pre-computed interpolation table is used to recover $p_{N_k}$ from $M_{\mathrm{max}}$ during the sampling.

We apply this change of parameters to both EoS models NL (linear) and
NP (polytropic), and refer to the variant models by ML and MP,
respectively. Similar to the choice of EoS models above, we make the prior choice that the NL and NP 
models have the same weight as the ML and MP models, respectively.

\subsection{Computational Details}
The Bayesian inference is performed via Markov Chain Monte Carlo (MCMC) using affine-invariant 
sampling, which explores the parameter space using an ensemble of ``walkers''. This sampling algorithm 
reduces the autocorrelation times, particularly when the target distribution has no complicated geometry. 
We use more than five times as many walkers as parameters. However, the cost of affine-invariant sampling 
scales with the number of parameters, resulting in a long equilibration time for the ensemble of walkers.

To assess the convergence of the MCMC chains, we use the Gelman-Rubin diagnostic which compares the 
variance between multiple chains to the variance within each chain. A value close to 1 indicates good 
convergence. In addition, we compute the average log-likelihood over the walkers and verify that in any 
given run, the first half of the log-likelihood trace matches or nearly overlaps with the second half. 
Due to our high-dimensional parameter space, the latter method proved to be more robust in diagnosing 
convergence and therefore used as the stopping criterion.

Once the MCMC convergence is achieved, we continue sampling to obtain an additional $\sim$ 500,000 samples at equilibrium. We discard the first $\sim 90$\% of all samples as burn-in and combine the remaining samples to compute the median autocorrelation time for each parameter over the walkers. Finally, we thin the chains by the maximum median autocorrelation time of all parameters to ensure that the samples are effectively independent.

\section{Results and Discussions \label{sec:rd}}
In this section, we present our results and compare them with previous works. For reference, models NL and 
NP refer to the linear and polytropic EoS, respectively. Models ML and MP are the variants of NL and NP, 
respectively, each with a uniform prior assumption on the maximum mass.

\subsection{Inferred EoS}
\begin{figure}[htbp]
	\centering
	\includegraphics[scale=0.5]{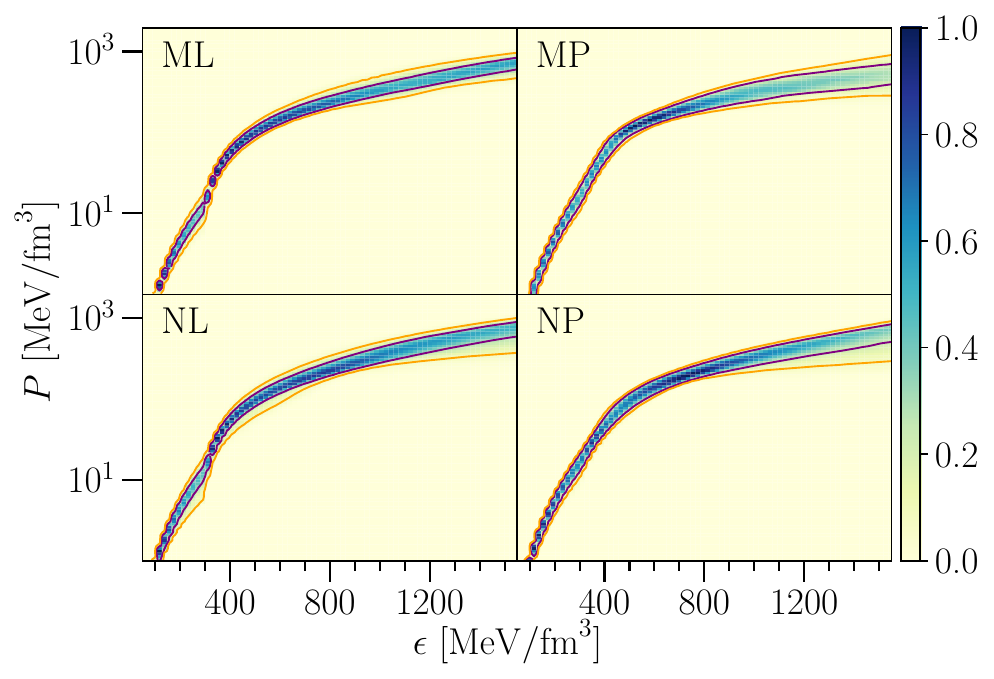}
	\caption{EoS models with $68\%$ (purple) and $95\%$ (orange) confidence levels. The color map shows locally normalized densities.}
	\label{fig:eos}
\end{figure}

Figure \ref{fig:eos} shows the posteriors of the pressure as a function of the energy density. The models ML 
and MP have nearly symmetric 68\% (purple) and 95\% (orange) confidence levels around the median lines, 
while NL and NP are skewed toward lower pressures at energy densities $\epsilon \ge$ 800 MeV/fm$^3$. 
The uncertainties generally increase with the density for all models, particularly beyond 1000 MeV/fm$^3$, 
indicating a strong dependence on the prior choices. The central energy densities are located between 
1100$-$1300 MeV/fm${^3}$ for the maximum mass star.

The models MP and NP have lower pressures at low energy densities and this behavior remains unchanged at 
high densities, which results into relatively lower pressures near the central energy density than ML and 
NL. Another characteristic feature of ML and NL is the sharp increase in pressure at $\epsilon \gtrsim$ 300 
MeV/fm$^3$, where the derivative $dP/d\epsilon=c_s^2$ is discontinuous, which is a manifestation of the 
prior choice for either of these models.

\begin{figure}[htbp]
	\centering
	\includegraphics[scale=0.5]{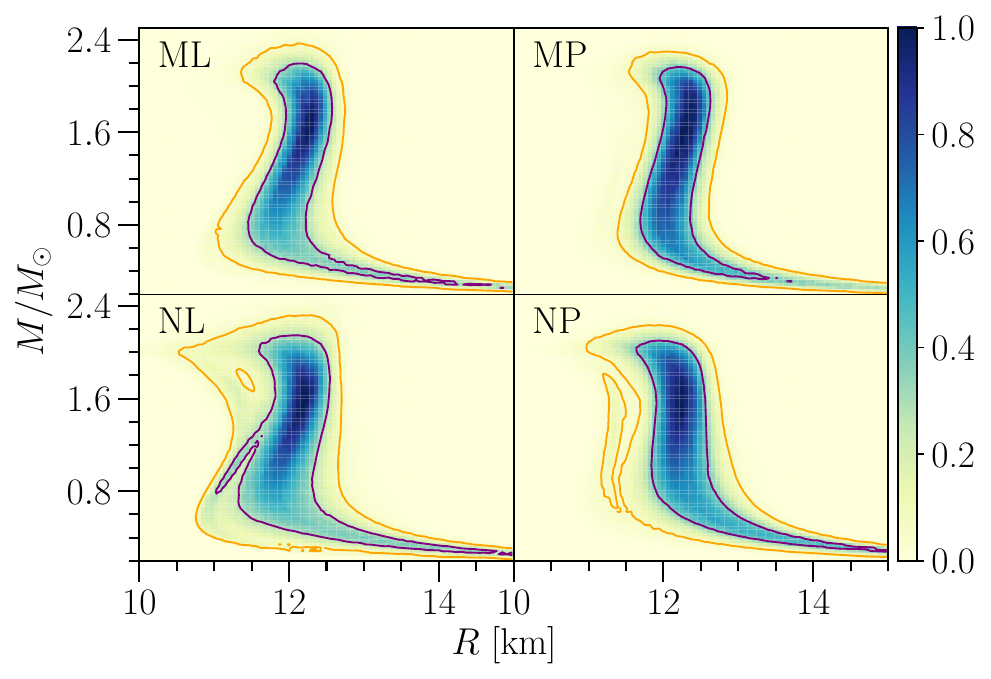}
	\caption{Mass-radius curves for the corresponding EoS in Figure \ref{fig:eos}, with $68\%$ (purple) and 
		$95\%$ (orange) confidence levels. The color map shows locally normalized densities.}
	\label{fig:mvsr}
\end{figure}

Figure \ref{fig:mvsr} shows the posterior of the NS radius as a function of the gravitational mass for each 
EoS. The mass-radius curves for ML and NL indicate smaller radii for low-mass stars and larger radii for 
high-mass stars. While MP closely follows this trend with a tighter lower bound for low-mass stars, NP 
exhibits a rather neutral behavior where low- and high-mass stars have fairly similar radii. The peak in 
each curve represents the maximum mass star. Generally, the curves in ML and MP are better constrained 
due to a different prior on $\mmax$. The artifacts seen in the bottom two panels (NL and NP) are 
noise arising from computing densities with sparse data points.

\begin{figure}[htbp]
	\centering
	\includegraphics[scale=0.5]{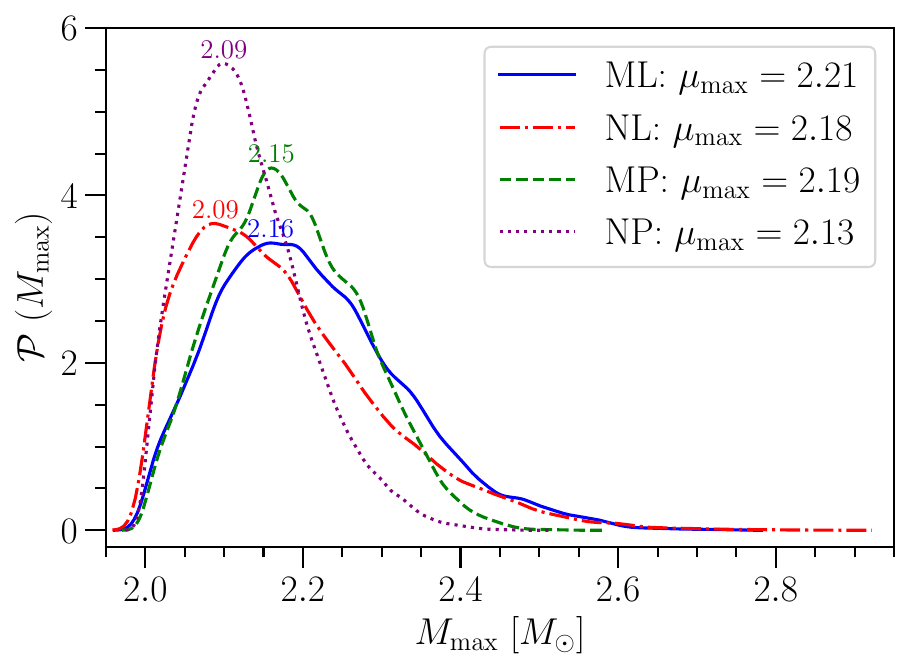}
	\caption{The posterior distributions of the NS maximum mass. The maximum a posteriori (MAP) values 
	are shown at the peaks and the means ($\mu_{\rm max}$) in the inset.}
	\label{fig:mmax}
\end{figure}

Figure \ref{fig:mmax} shows posteriors of the maximum mass for each model. The cutoff at 2$\M$ reflects our 
imposed constraint on the lower bound that $\mmax \ge 2\M$. Comparing NL with ML (and NP with MP), we see 
that prior assumption on $\mmax$ shifts the peak and the upper bound to favor larger $\mmax$, thus 
supporting more massive stars. The model NL accommodates higher $\mmax$ with its widest distribution and 
support up to 2.9$\M$. Conversely, NP has the narrowest width and support with its peak close to NL. A 
similar argument can be made for the pair ML and MP. The difference in these distribution shapes suggests 
that EoS prior choices impact their widths.

Table \ref{tab:res} contains our results for the radius and tidal deformability for a $1.4\M$ star 
($R_{1.4}$ and $\Lambda_{1.4}$), the radius and mass of the maximum-mass star ($\rmax$ and $\mmax$), for
each EoS model, along with the results from most recent works. We present all results at 90\% confidence 
level a more direct comparison with previous works. Our posteriors for $\mmax$ are smaller than other works 
in general, particularly NL and NP, but remain within bounds at 90\% confidence level. This discrepancy can 
be attributed to different EoS parametrizations as well as the data sets included in the previous works. For 
instance, \cite{fan-24} clearly includes a larger number of massive pulsars ($\ge 1.9 \M$) than this work. 
However, they also demonstrate how exclusion of a few such NSs significantly shifts the $\mmax$ posterior 
toward the low-mass end ($\sim 2.16 \M$, see Figure 9 in this reference), which is consistent with this work.
Our posteriors for $\rmax$ in Table \ref{tab:res} shows that the models NL and NP favor relatively smaller 
radii with larger uncertainties, compared to ML and MP, respectively. The skewness in the confidence 
intervals can be attributed to the peaks of the mass-radius curves in Figure \ref{fig:mvsr}, by drawing 
horizontal lines near the maximum mass regions. The overall narrower bounds for $\rmax$ in ML and MP 
indicates that different prior choices on $\mmax$ imposes further constraints on $\rmax$. Our results for 
$\rmax$ are smaller than \cite{fan-24}. However, due to the positive correlation between $\mmax$ and $\rmax$ 
(see Figure 3 in this reference), a similar argument can be made that exclusion of a few massive pulsars 
would reduce $\rmax$ closer to our results.

\begin{deluxetable}{lcccc}[htbp]
	\tablecaption{Comparison of our results with most recent works (a)--(c) at 90\% confidence level, where 
		a blank spot (-) indicates that the result was not reported in the reference. \label{tab:res}}
	\tablehead{
		\colhead{Ref.} &
		\colhead{$M_\mathrm{max}$ [$M_\odot$]} &
		\colhead{$R_\mathrm{max}$ [km]} &
		\colhead{$R_{1.4}$ [km]} &
		\colhead{$\Lambda_{1.4}$}
	}
	\startdata
	ML & $2.16_{-0.12}^{+0.27}$ & $11.45_{-0.61}^{+0.56}$ & $12.12_{-0.39}^{+0.39}$ & $438_{-84}^{+47}$ \\
	NL & $2.09_{-0.07}^{+0.33}$ & $11.33_{-1.26}^{+0.66}$ & $12.09_{-0.70}^{+0.41}$ & $415_{-143}^{+64}$ \\
	MP & $2.15_{-0.10}^{+0.19}$ & $11.76_{-0.68}^{+0.41}$ & $12.18_{-0.42}^{+0.35}$ & $438_{-97}^{+66}$ \\
	NP & $2.09_{-0.07}^{+0.18}$ & $11.28_{-0.77}^{+0.78}$ & $12.24_{-0.69}^{+0.36}$ & $415_{-121}^{+68}$ \\
	(a) & $2.25_{-0.11}^{+0.16}$ & $11.90_{-0.94}^{+1.05}$ & - & - \\
	(b) & $2.22_{-0.19}^{+0.21}$ & - & $12.34_{-0.53}^{+0.43}$ & $436_{-117}^{+109}$ \\
	(c) & $2.28_{-0.21}^{+0.41}$ & - & $12.2_{-0.9}^{+0.8}$ & $438_{-166}^{+224}$ \\
	\enddata
	\tablecomments{(a) \cite{fan-24}, (b) \cite{biswas-25a}, (c) \cite{golomb-25}}
\end{deluxetable}

Next, our results for $R_{1.4}$ show that the EoS priors only impact the widths of the posteriors while the 
modes are mostly unaffected. For example, the 90\% confidence intervals are wider for models NL and NP, 
closely matching those in \cite{biswas-25a}, but are narrower than \cite{golomb-25}. Our central values of 
$R_{1.4}$ are generally smaller than both works, particularly compared to \cite{biswas-25a}, who obtained 
slightly larger $R_{1.4}$ due to information from $\chi$EFT. However, our results are fairly close and 
remain within their uncertainty bounds.

\begin{figure}[htbp]
	\centering
	\includegraphics[scale=0.5]{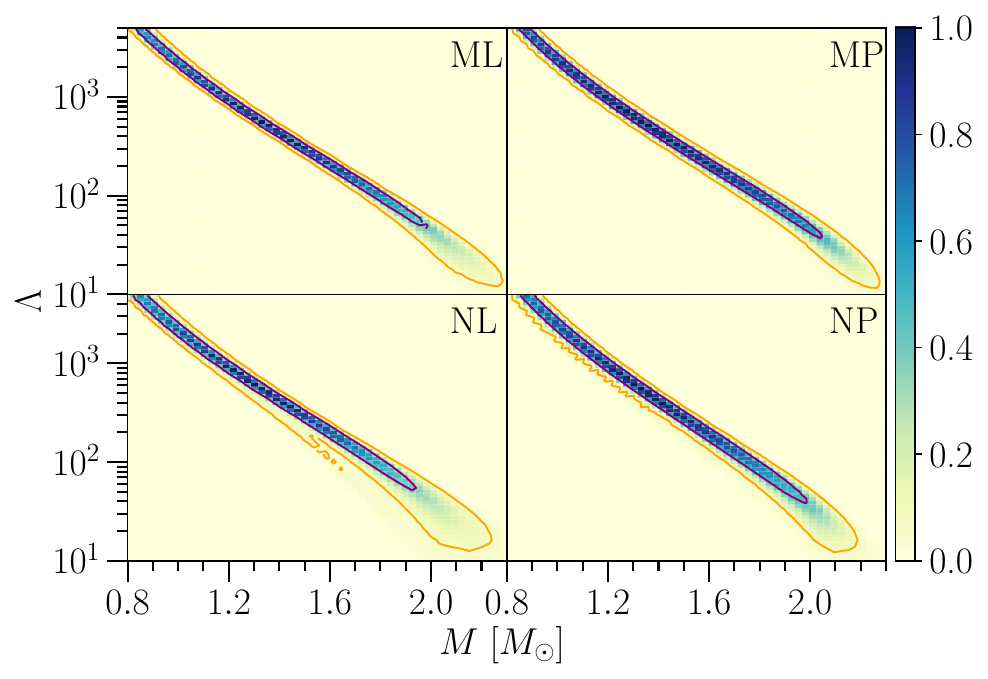}
	\caption{Tidal deformability as a function of NS mass for each model, with $68\%$ (purple) and $95\%$ 
		(orange) confidence levels. The color map shows locally normalized densities.}
	\label{fig:lambda-m}
\end{figure}

Figure \ref{fig:lambda-m} shows the dimensionless tidal deformability as a function of mass for each model. 
We deliberately limit the low-mass ends in the figure to keep the tides within physically meaningful values. 
The tidal deformabilities of a typical 1.4$\M$ star ($\Lambda_{1.4}$) for different EoS are reported in 
Table \ref{tab:res}. All models constrain the $\Lambda-M$ curves well up to $\sim 1.9 \M$, beyond which they 
become dominated by large uncertainties. However, comparing the ML, MP (top row) with NL, NP (bottom row) 
suggests that prior knowledge of $\mmax$ results into a tighter distribution of $\Lambda(M)$, particularly 
for massive NSs. Our results for $\Lambda_{1.4}$, particular for models ML and MP, are consistent with 
recent works (see Table \ref{tab:res}) with narrower bounds at 90\% confidence level. 

\subsection{Inferred Mass Distributions}
As explained in Subsection \ref{sub:mdist}, the uncertainties in mass measurements are handled by
the individual mass parameters, and consequently, are not explicitly marginalized over in Equation 
(\ref{eq:wgt}). We verified that the posterior distributions of all mass parameters in all EoS models are 
fairly consistent within the statistical uncertainties of the measured masses.

\begin{figure*}[htbp]
	\centering
	\includegraphics[scale=0.5]{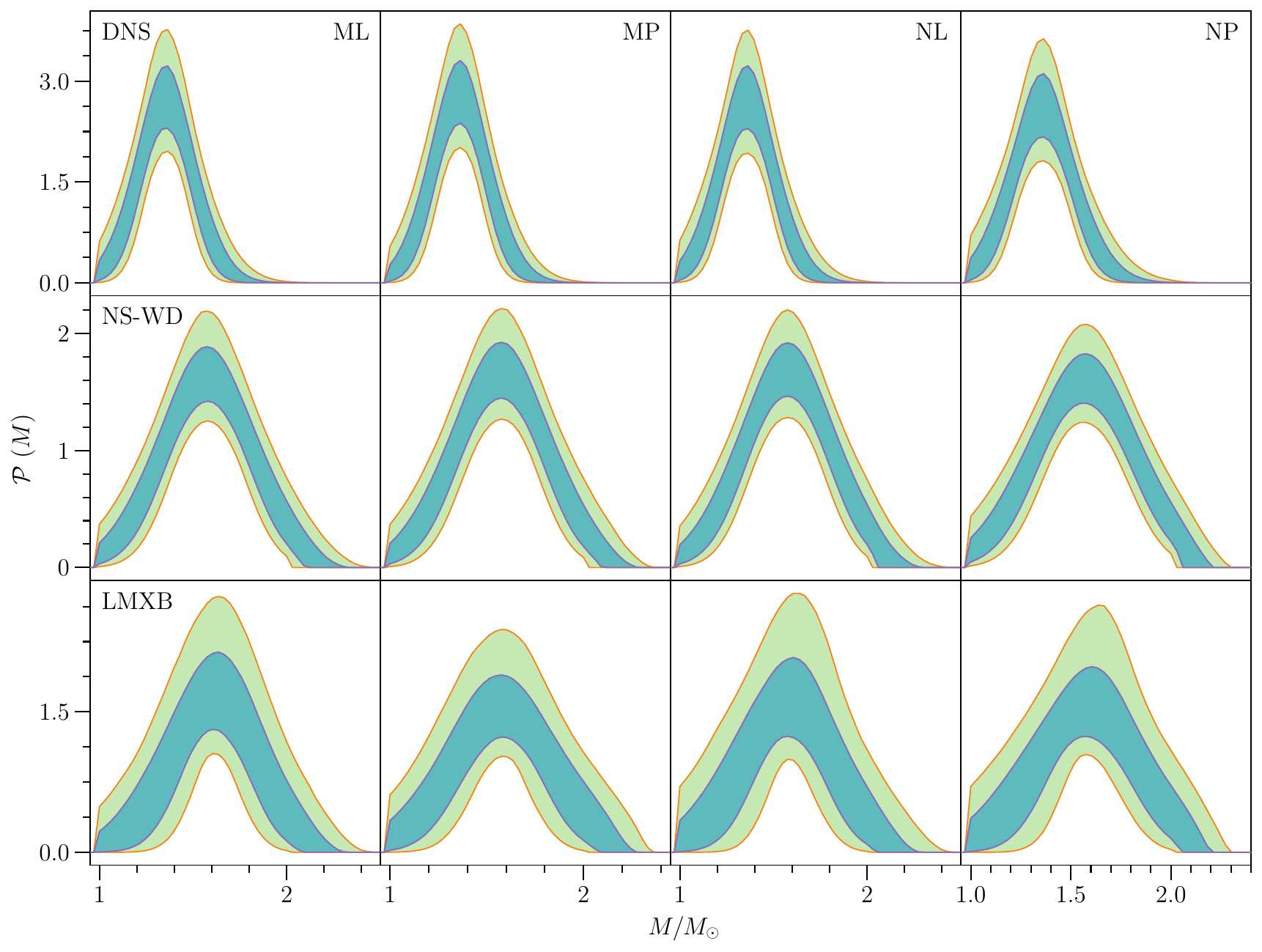}
	\caption{The normalized mass distributions of NS populations (row-wise), grouped by the EoS models 
		(column-wise). The contour lines represent $68\%$ (purple) and $95\%$ (orange) confidence levels. 
		The statistics are reported in Table \ref{tab:dist}.}
	\label{fig:dist}
\end{figure*}

Figure \ref{fig:dist} shows the posterior mass distributions of the populations grouped by the EoS models. 
Each column represents the same EoS and each row shows the mass distributions of a given population. The 
shapes of the DNS and NS-WD distributions do not deviate significantly across different EoS models, because 
all the stars (except GW170817 in DNS) have mass measurements with no information of their radii. Thus, 
their only connection to the EoS is through the maximum mass constraint. On the other hand, the LMXB 
distributions suggest a stronger dependence on the EoS because most stars in LMXB have mass-radius 
observations (see Table \ref{tab:lmxb}). Therefore, the EoS priors strongly influence the masses through 
their radii, along with the maximum mass constraint. The sharper truncations at the high-mass tails 
of the NS-WD and LMXB mass distributions for MP and NP stem from the narrower supports for $\mmax$ than for 
ML and NL in Figure \ref{fig:mmax}.

Table \ref{tab:dist} contains statistics of the mass distributions in Figure \ref{fig:dist} showing the 
posterior shape parameters: mean, width, and skewness for each population and EoS model at 68\% 
confidence level. The smaller population means in NP, particularly for LMXB, may be related to the 
narrower $\mmax$ support for NP (see Figure \ref{fig:mmax}). The population widths ($\sigma$) do not 
drastically change for different EoS models because they are primarily driven by the dispersion in 
observed masses of each type. While the population skewness ($\alpha$) parameters do nominally change 
across the different EoS model families and priors, these changes are comparable to the statistical 
uncertainty in this parameter. Note that MP and NP produce nearly identical NS-WD and LMXB 
distributions except for their skewnesses.

\begin{deluxetable}{l l c c c}[t] \label{tab:dist}
	\tablecaption{Maximum a posteriori (MAP) at 68\% confidence level for the shape parameters mean ($\mu$), 
		width ($\sigma$), and skewness ($\alpha$) of the mass distributions in Figure \ref{fig:dist}.}
	\tablehead{
		\colhead{Model} & \colhead{Parameter} & \colhead{DNS} & \colhead{NS-WD} & \colhead{LMXB}
	}
	\startdata
	& $\mu$ [$M_\odot$]   & $1.39_{-0.07}^{+0.04}$ & $1.50_{-0.06}^{+0.14}$ & $1.71_{-0.15}^{+0.08}$ \\
	ML & $\sigma$ [$M_\odot$]& $0.14_{-0.01}^{+0.04}$ & $0.26_{-0.04}^{+0.05}$ & $0.24_{-0.06}^{+0.07}$ \\
	& $\alpha$            & $-0.25_{-0.35}^{+0.48}$ & $0.75_{-1.03}^{+0.00}$ & $-0.48_{-0.25}^{+0.51}$ \\
	\hline
	& $\mu$ [$M_\odot$]   & $1.41_{-0.08}^{+0.03}$ & $1.53_{-0.09}^{+0.13}$ & $1.62_{-0.11}^{+0.16}$ \\
	NL & $\sigma$ [$M_\odot$]& $0.15_{-0.02}^{+0.04}$ & $0.24_{-0.02}^{+0.06}$ & $0.25_{-0.06}^{+0.09}$ \\
	& $\alpha$            & $-0.49_{-0.27}^{+0.65}$ & $0.83_{-1.27}^{+0.07}$ & $-0.93_{-0.10}^{+1.04}$ \\
	\hline
	& $\mu$ [$M_\odot$]   & $1.38_{-0.06}^{+0.05}$ & $1.61_{-0.14}^{+0.08}$ & $1.62_{-0.12}^{+0.15}$ \\
	MP & $\sigma$ [$M_\odot$]& $0.13_{-0.01}^{+0.05}$ & $0.24_{-0.03}^{+0.06}$ & $0.24_{-0.03}^{+0.11}$ \\
	& $\alpha$            & $-0.30_{-0.26}^{+0.65}$ & $0.03_{-0.59}^{+0.49}$ & $-0.41_{-0.32}^{+0.57}$ \\
	\hline
	& $\mu$ [$M_\odot$]   & $1.30_{-0.03}^{+0.09}$ & $1.53_{-0.10}^{+0.12}$ & $1.51_{-0.13}^{+0.13}$ \\
	NP & $\sigma$ [$M_\odot$]& $0.14_{-0.01}^{+0.06}$ & $0.26_{-0.03}^{+0.06}$ & $0.24_{-0.04}^{+0.12}$ \\
	& $\alpha$            & $0.38_{-0.56}^{+0.40}$ & $0.23_{-0.56}^{+0.45}$ & $0.92_{-0.99}^{+0.07}$ \\
	\enddata
\end{deluxetable}

\subsubsection{Double Neutron Stars}
The NS mass distribution for members of DNS is well-constrained by the many precise NS mass measurements 
enabled by pulsar timing. Both the population mean $\mu_{\rm dns}$ and population width $\sigma_{\rm dns}$ 
are extremely well constrained by this mass information alone. Conversely, we anticipate these DNS 
observations provide relatively little direct information about the EoS, since few have radius information 
and none are close to the NS maximum mass. 

For direct comparison of the NS mass distributions, we present all results from this and previous works 
within 68\% confidence levels. Our results for the DNS population are very close to the population mean
$\mu_{\rm dns}=1.33\M$ identified in previous work \citep{ozel-12, kiziltan-13}, within the statistical 
uncertainties. Our inferred population dispersion, however, is somewhat larger than previously stated 
inferences for DNS, i.e., $\sigma_{\rm dns}=0.05\M$ from previous work \citep{ozel-12, kiziltan-13}, because 
our sample includes more NS with a wider range of masses, many of which were not available in these earlier 
analyses; see Table \ref{tab:dns}.

%\editremark{offline: please check F. Ozel's paper, as to why the event list is so much smaller.}

\subsubsection{Low-Mass X-ray Binaries}
For NS hosted in LMXB, however, the inferred population is substantially different. As inputs, these LMXB 
observations extend to significantly higher masses, approaching the NS maximum mass, but are much less 
numerous and much less precisely determined than the DNS case. However, these observations often also 
provide us with invaluable information about NS radii. As in previous studies of LMXB, the NS mass 
measurements in this range allow us to infer a characteristic mass $\mu$ and width $\sigma$. Our inferred 
values are again similar to results obtained in previous studies, keeping in mind most other
analyses use multimodal LMXB mass models. 

Our model NP has a similar peak near $\mu_{\rm lmxb}=1.48\M$, $\sigma_{\rm lmxb}=0.20\M$ 
\citep{ozel-12} for recycled NSs, whereas other models have peaks at slightly larger masses. Due to most of 
the recent works using a bimodal Gaussian mass distribution for all NSs as a single population, it is 
difficult to directly compare our results with them. For example, \cite{fan-24} and \cite{biswas-25a} both 
obtained the first peak around $1.33-1.34\M$ with a narrow width of $0.08\M$, and a wider second peak around 
$1.67-1.74\M$ with widths between $0.22-0.24\M$. Our DNS results are similar to the first and LMXB to the 
second peak, while NS-WD falls between them.
Unlike most other analyses however, our investigation self-consistently deduces the NS EoS from these and 
other observations and hence imposes a fundamental maximum mass. Because we use NS radius information, not 
merely mass measurements, our inferred NS mass distribution has a sharp upper limit, imposed by the EoS.
Previous investigations have attempted a joint analysis with many NS mass and radius information to 
constraint the NS in a qualitatively similar manner, notably \cite{fan-24, biswas-25a}; however, these 
analyses differ in multiple respects (e.g., using a single NS population for all events; the EoS models 
employed) and arrives at a systematically much larger maximum NS mass compared to some of our analyses. 

\subsubsection{Neutron Star -- White Dwarf Binaries}
For NS hosted in WD, the inferred population lies in between the two extremes above. As inputs, these NS-WD 
binaries have more poorly constrained masses than NS in DNS, but provide some additional information about 
NS radii. Like NS in LMXB, their masses also extend to the NS maximum mass.
For NS-WD, our results are consistent with the peak reported in \cite{ozel-12} for recycled NSs, with a 
population mean of $1.46\M$ or $1.55\M$ in \cite{kiziltan-13}, and a population dispersion of $0.21\M$
in both works.

\begin{figure}[htbp]
	\centering
	\includegraphics[scale=0.5]{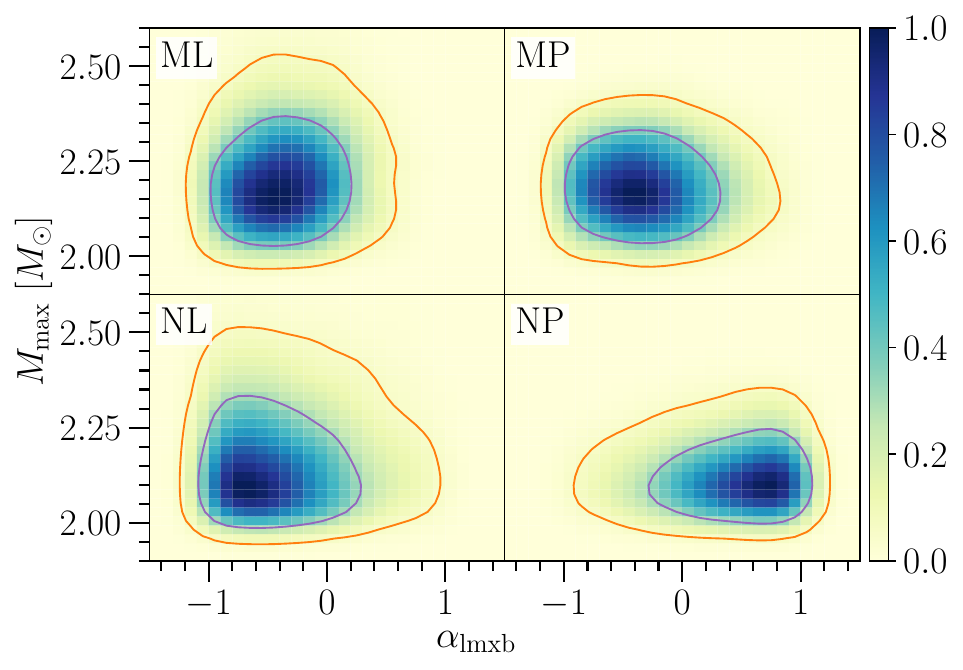}
	\caption{Correlation between the maximum mass and the skewness of the LMXB mass distribution for each 
		EoS model, with $68\%$ (purple) and $95\%$ (orange) confidence levels.}
	\label{fig:mmax-alpha_lmxb}
\end{figure}

\subsubsection{Correlations between EoS and Mass Distributions}
Finally, we examine the correlations between the mass distribution parameters (mean, width, skewness) and 
the maximum mass. We observe no significant correlations except for the skewness parameter of the 
LMXB distribution. Figure \ref{fig:mmax-alpha_lmxb} shows the skewness is uncorrelated with the maximum 
mass for models ML and MP. However, there is a negative and positive correlation between them for NL and 
NP, respectively.

\section{Conclusions \label{sec:con}}
In this work, we perform joint inference of the nuclear EoS and the mass distribution of NS present in three 
different astrophysical populations: double NS (DNS), NS-WD, and LMXB. For NS populations, we find results 
consistent with previous investigations, but critically now supplemented for the NS-WD and LMXB populations 
by a maximum mass imposed by the nuclear EoS. As seen previously, we find distinct mass distributions for 
each component, and no significant correlations between the shape parameters (mean, width, skewness) across 
the DNS, NS-WD, and LMXB populations. For EoS inference, we find a maximum-mass NS which is qualitatively 
consistent with previous investigations: between $2.0-2.5 \M$ (union over methods). The specific 
quantitative result, however, depends critically on how we propagate prior knowledge about observables into 
our inference: using a uniform prior on $\mmax$ explicitly can shift the posterior $\mmax$ toward the 
high-mass end than a prior imposed solely on the EoS parametrization itself. 

Although the mass distribution function in Equation (\ref{eq:sn}) is defined over the entire real line,
our combined likelihood (Equation \ref{eq:wgt}) is nonzero only over the NS mass range $[\mmin, \mmax]$. 
Since $\mmin$ is fixed at $1\M$, the normalization constant in Equation (\ref{eq:sn}) should explicitly depend 
on $\mmax$ when marginalized over the mass range, which can influence the likelihood and, consequently, the 
posteriors and potential correlations involving $\mmax$. However, we tested this dependence and found that 
the normalization constant is nearly flat against $\mmax$. Additionally, we reproduced the posterior $\mmax$ 
distributions and compared with Figure \ref{fig:mmax}. We observed that the means ($\mu_\mathrm{max}$) are 
shifted upward by $\sim 0.001\M$ when the normalization constant is accounted for, while the maximum a 
posteriori (MAP) values remained identical up to three decimal places. We further examined the correlations 
between $\mmax$ and the shape parameters of the mass distributions for all populations and EoS models, and 
found that normalization does not change the correlations either. Therefore, our analysis does not implement 
the normalization constant since doing so would not significantly impact our likelihood and posterior 
results.

Our method contrasts previous studies that treated $\mmax$ as a truncation parameter to infer 
EoS-insensitive mass distributions driven by observational data \citep{antoniadis-16, shao-20}, including 
works that went further by using the inferred $\mmax$ to characterize EoS models \citep{alsing-18, fan-24}. 
However, as recent discoveries of more massive pulsars continue to increase the lower bound on $\mmax$, the 
upper bound remains elusive. Recently, joint inference of the EoS and NS mass distribution has become a 
predominantly common choice \citep{wysocki-20, landry-21, golomb-22, biswas-25a, golomb-25}. With that 
shared goal, this work further shows that changing the prior distribution of the observable $\mmax$ itself, 
rather than that of a nuisance EoS parameter which is unpredictable, results into a different $\mmax$ 
posterior with a larger upper bound. 

Following \citet{golomb-25}, one could introduce an astrophysical maximum mass $M_\mathrm{pop}$ for 
each NS population and examine its relation to $\mmax$. We do not include such an additional parameter here, 
but our results already show the relevant behavior. For DNS systems, the high-mass tails lie well below 
$\mmax$ (see Figure~\ref{fig:mmax}) because the measured DNS masses do not extend to the $\mmax$ cutoff. 
An inferred $M_\mathrm{pop}<\mmax$ for DNS would therefore be controlled by the most massive observed DNSs 
rather than by the theoretical maximum mass. In contrast, the NS-WD and LMXB distributions extend to higher 
masses and are truncated near $M_\mathrm{pop}\approx\mmax$. Thus, our analysis demonstrates this 
population-dependent cutoff behavior without introducing a separate $M_\mathrm{pop}$ parameter for each 
population.

Because we separate each NS population into its own distinct category, rather than assume a universal NS mass
distribution, we avoid numerous subtleties associated with selection effects and evolutionary differences 
between each of these categories: reconstructing the universal NS mass distribution from observed compact 
binaries is quite difficult. As a result, our approach is inherently robust and makes observationally 
pertinent predictions.  However, we do make some simplifying assumptions which could be in error.  For 
example, we assume that DNS detected by GW observations are presumed to have the same mass distribution as 
other galactic DNS binaries. In practice, the relatively small contamination of these events in a relatively 
large sample suggests that even a misclassification would not significantly change our overall results. 
Conversely, only two DNS observations are insufficient to classify GWs as a different population, assuming 
that GW190425 is indeed a DNS. Additionally, we assume NSs within each category are drawn from the same 
underlying distribution, such that the categorization is meaningful (e.g., NS with the same companion types 
but are in different evolution stages or have accretion histories also follow the same underlying 
distribution). 

For simplicity, in our analysis we followed the approach of \cite{kiziltan-13}, using a single mass 
distribution with skewness for each component. By contrast, many previous works adopt more complex bimodal 
mass distributions in order to explain all NS categories in one population. As noted above, we do not 
believe the observational and evolution selection effects can be inverted to allow us to recover the  
overall underlying NS population. Additionally, for each component, we prefer to retain the simple models 
adopted here, since the large uncertainty bands of NS mass measurements in NS-WD and LMXB would also hinder 
resolving multiple peaks even if they exist. 

We conclude by discussing the scope and limitations of our work. First, despite our best efforst to keep
our data set up to date, by the time we noticed the updated meaurements of J0030+0451 \citep{vinciguerra-24}
and J0740+6620 \citep{salmi-24}, our MCMC simulations had progressed toward convergence. Changing the data at
that stage would result into a longer equilibration time and hence we decided to keep the original data for 
these stars. Second, we refrain from discussing possible systematic biases in radius measurements in this work 
as this issue has been covered in detail by \citet{chatziioannou-25}. Finally, by limiting our analysis to 
DNS, NS-WD, and LMXB, we exclude other populations such as NS--black holes, high-mass X-ray binaries, and 
isolated neutron stars from our analysis. Even within the populations included here, we exclude 
binaries where individual NS masses are not directly measured. Thus, the selected NS binaries may not be 
representatives of the entire NS populations. Moreover, our inferred mass distributions is also biased by the 
inevitable selection effects such as GW detection being more sensitive to massive binaries, optical 
observations being more favorable for compact objects etc. Therefore, future multi-messenger observations of 
more neutron stars may place tighter constraints on the dense matter equation of state and the maximum mass 
and bring us closer to revealing the underlying NS mass distributions.

\begin{acknowledgments}
We thank Satyajit Roy, Sanket Sharma, and Zidu Lin for their useful insights and help in developing the code, and for raising thoughtful questions. This project is supported by the National Science Foundation grant \href{https://www.nsf.gov/awardsearch/showAward?AWD_ID=2206322}{AST 22-06322}.
\end{acknowledgments}

\section*{Data Availability}
The data files are publicly available on Zenodo \dataset[doi:10.5281/zenodo.17842420]
{https://doi.org/10.5281/zenodo.17842420} \citep{anik-25}. We used the C++ code \texttt{BAMR} \citep{bamr} and its dependency \texttt{O2SCL} \citep{o2scl} to perform Markov Chain Monte Carlo simulations, which are also public and can be found on \href{https://github.com/awsteiner/o2scl/tree/hmc}{Github (\texttt{awsteiner/o2scl})}.

\bibliography{aastex701}

@misc{bamr,
	author        = {{Steiner}, Andrew W.},
	title         = "{BAMR: Bayesian Analysis of Mass and Radius Observations}",
	howpublished  = {Astrophysics Source Code Library, record ascl:1408.020},
	year          = 2014,
	month         = aug,
	eid           = {ascl:1408.020},
	adsurl        = {https://ui.adsabs.harvard.edu/abs/2014ascl.soft08020S}
}

@misc{o2scl,
	author       = {{Steiner}, Andrew W.},
	title        = "{O2scl: Object-oriented Scientific Computing Library}",
	howpublished = {Astrophysics Source Code Library, record ascl:1408.019},
	year         = 2014,
	month        = aug,
	eid          = {ascl:1408.019},
	adsurl       = {https://ui.adsabs.harvard.edu/abs/2014ascl.soft08019S}
}

@misc{anik-25,
	author       = {Anik, Mahmudul Hasan},
	title        = {Inference of Neutron Star Mass Distributions and the Equation of State from
					Multi-messenger Observations},
	month        = dec,
	year         = 2025,
	publisher    = {Zenodo},
	version      = {1.0.0},
	doi          = {10.5281/zenodo.17842420},
	url          = {https://doi.org/10.5281/zenodo.17842420},
}

@article{vinciguerra-24,
doi = {10.3847/1538-4357/acfb83},
url = {https://doi.org/10.3847/1538-4357/acfb83},
year = {2024},
month = {jan},
publisher = {The American Astronomical Society},
volume = {961},
number = {1},
pages = {62},
author = {Vinciguerra, Serena and Salmi, Tuomo and Watts, Anna L. and Choudhury, Devarshi and Riley, Thomas E. and Ray, Paul S. and Bogdanov, Slavko and Kini, Yves and Guillot, Sebastien and Chakrabarty, Deepto and Ho, Wynn C. G. and Huppenkothen, Daniela and Morsink, Sharon M. and Wadiasingh, Zorawar and Wolff, Michael T.},
title = {An Updated Mass-Radius Analysis of the 2017-2018 NICER Data Set of PSR J0030+0451},
journal = {The Astrophysical Journal},
}

@article{salmi-24,
doi = {10.3847/1538-4357/ad5f1f},
url = {https://doi.org/10.3847/1538-4357/ad5f1f},
year = {2024},
month = {oct},
publisher = {The American Astronomical Society},
volume = {974},
number = {2},
pages = {294},
author = {Salmi, Tuomo and Choudhury, Devarshi and Kini, Yves and Riley, Thomas E. and Vinciguerra, Serena and Watts, Anna L. and Wolff, Michael T. and Arzoumanian, Zaven and Bogdanov, Slavko and Chakrabarty, Deepto and Gendreau, Keith and Guillot, Sebastien and Ho, Wynn C. G. and Huppenkothen, Daniela and Ludlam, Renee M. and Morsink, Sharon M. and Ray, Paul S.},
title = {The Radius of the High-mass Pulsar PSR J0740+6620 with 3.6 yr of NICER Data},
journal = {The Astrophysical Journal},
}

@article{chatziioannou-25,
  title = {Neutron stars and the dense matter equation of state},
  author = {Chatziioannou, Katerina and Cromartie, H. Thankful and Gandolfi, Stefano and Tews, Ingo and Radice, David and Steiner, Andrew W. and Watts, Anna L.},
  journal = {Rev. Mod. Phys.},
  volume = {97},
  issue = {4},
  pages = {045007},
  numpages = {49},
  year = {2025},
  month = {Dec},
  publisher = {American Physical Society},
  doi = {10.1103/ymsq-cfcw},
  url = {https://link.aps.org/doi/10.1103/ymsq-cfcw}
}

@article{lange-18,
	author = {{Lange}, Jacob and {O'Shaughnessy}, Richard and {Rizzo}, Monica},
	title = "{Rapid and accurate parameter inference for coalescing, precessing compact binaries}",
	journal = {arXiv e-prints},
	year = 2018,
	month = may,
	eid = {arXiv:1805.10457},
	pages = {arXiv:1805.10457},
	doi = {10.48550/arXiv.1805.10457},
	archivePrefix = {arXiv},
	eprint = {1805.10457},
	primaryClass = {gr-qc},
	adsurl = {https://ui.adsabs.harvard.edu/abs/2018arXiv180510457L}
}

@article{adhikari-22,
	title = {Precision Determination of the Neutral Weak Form Factor of $^{48}\mathrm{Ca}$},
	author={Adhikari, D and Albataineh, H and Androic, D and Aniol, KA and Armstrong, DS and Averett, T and Ayerbe Gayoso, C and Barcus, SK and Bellini, V and Beminiwattha, RS and others},
	collaboration = {CREX Collaboration},
	journal = {Phys. Rev. Lett.},
	volume = {129},
	issue = {4},
	pages = {042501},
	numpages = {8},
	year = {2022},
	month = {Jul},
	publisher = {American Physical Society},
	doi = {10.1103/PhysRevLett.129.042501},
	url = {https://link.aps.org/doi/10.1103/PhysRevLett.129.042501}
}

@article{adhikari-21,
	title = {Accurate Determination of the Neutron Skin Thickness of $^{208}\mathrm{Pb}$ through Parity-Violation in Electron Scattering},
	author={Adhikari, D and Albataineh, H and Androic, D and Aniol, K and Armstrong, DS and Averett, T and Ayerbe Gayoso, C and Barcus, S and Bellini, V and Beminiwattha, RS and others},
	collaboration = {PREX Collaboration},
	journal = {Phys. Rev. Lett.},
	volume = {126},
	issue = {17},
	pages = {172502},
	numpages = {7},
	year = {2021},
	month = {Apr},
	publisher = {American Physical Society},
	doi = {10.1103/PhysRevLett.126.172502},
	url = {https://link.aps.org/doi/10.1103/PhysRevLett.126.172502}
}

@article{tews-25,
	title={Neutron matter from local chiral effective field theory interactions at large cutoffs},
	volume={7},
	ISSN={2643-1564},
	url={http://dx.doi.org/10.1103/r314-6r62},
	DOI={10.1103/r314-6r62},
	number={3},
	journal={Physical Review Research},
	publisher={American Physical Society (APS)},
	author={Tews, Ingo and Somasundaram, Rahul and Lonardoni, Diego and Göttling, Hannah and Seutin, Rodric and Carlson, Joseph and Gandolfi, Stefano and Hebeler, Kai and Schwenk, Achim},
	year={2025},
	month=jul }

@article{schwab-10,
	doi = {10.1088/0004-637X/719/1/722},
	url = {https://doi.org/10.1088/0004-637X/719/1/722},
	year = {2010},
	month = {jul},
	publisher = {The American Astronomical Society},
	volume = {719},
	number = {1},
	pages = {722},
	author = {Schwab, J. and Podsiadlowski, Ph. and Rappaport, S.},
	title = {FURTHER EVIDENCE FOR THE BIMODAL DISTRIBUTION OF NEUTRON-STAR MASSES},
	journal = {The Astrophysical Journal}
}

@article{finn-94,
	title = {Observational Constraints on the Neutron Star Mass Distribution},
	author = {Finn, Lee Samuel},
	journal = {Phys. Rev. Lett.},
	volume = {73},
	issue = {14},
	pages = {1878--1881},
	numpages = {0},
	year = {1994},
	month = {Oct},
	publisher = {American Physical Society},
	doi = {10.1103/PhysRevLett.73.1878},
	url = {https://link.aps.org/doi/10.1103/PhysRevLett.73.1878}
}

@article{thorsett-99,
	doi = {10.1086/306742},
	url = {https://doi.org/10.1086/306742},
	year = {1999},
	month = {feb},
	publisher = {},
	volume = {512},
	number = {1},
	pages = {288},
	author = {Thorsett, S. E. and Chakrabarty, Deepto},
	title = {Neutron Star Mass Measurements. I. Radio Pulsars},
	journal = {The Astrophysical Journal}
}

@article{wysocki-20,
	title={Inferring the neutron star equation of state simultaneously with the population of merging neutron stars},
	author={Wysocki, Daniel and O'Shaughnessy, Richard and Wade, Leslie and Lange, Jacob},
	journal={arXiv preprint arXiv:2001.01747},
	year={2020}
}

@article{farr-20,
	doi = {10.3847/2515-5172/ab9088},
	url = {https://doi.org/10.3847/2515-5172/ab9088},
	year = {2020},
	month = {may},
	publisher = {The American Astronomical Society},
	volume = {4},
	number = {5},
	pages = {65},
	author = {Farr, Will M. and Chatziioannou, Katerina},
	title = {A Population-Informed Mass Estimate for Pulsar J0740+6620},
	journal = {Research Notes of the AAS},
}

@article{landry-21,
	doi = {10.3847/2041-8213/ac2f3e},
	url = {https://doi.org/10.3847/2041-8213/ac2f3e},
	year = {2021},
	month = {nov},
	publisher = {The American Astronomical Society},
	volume = {921},
	number = {2},
	pages = {L25},
	author = {Landry, Philippe and Read, Jocelyn S.},
	title = {The Mass Distribution of Neutron Stars in Gravitational-wave Binaries},
	journal = {The Astrophysical Journal Letters},
}

@article{li-21,
	doi = {10.3847/1538-4357/ac34f0},
	url = {https://doi.org/10.3847/1538-4357/ac34f0},
	year = {2021},
	month = {dec},
	publisher = {The American Astronomical Society},
	volume = {923},
	number = {1},
	pages = {97},
	author = {Li, Yin-Jie and Tang, Shao-Peng and Wang, Yuan-Zhu and Han, Ming-Zhe and Yuan, Qiang and Fan, Yi-Zhong and Wei, Da-Ming},
	title = {Population Properties of Neutron Stars in the Coalescing Compact Binaries},
	journal = {The Astrophysical Journal},
}

@article{golomb-22,
	doi = {10.3847/1538-4357/ac43bc},
	url = {https://doi.org/10.3847/1538-4357/ac43bc},
	year = {2022},
	month = {feb},
	publisher = {The American Astronomical Society},
	volume = {926},
	number = {1},
	pages = {79},
	author = {Golomb, Jacob and Talbot, Colm},
	title = {Hierarchical Inference of Binary Neutron Star Mass Distribution and Equation of State with Gravitational Waves},
	journal = {The Astrophysical Journal},
}

@article{fan-24,
	title = {Maximum gravitational mass $M_{\mathrm{TOV}} = 2.25_{-0.07}^{+0.08}\,M_{\odot}$ inferred at about 3\% precision with multimessenger data of neutron stars},
	author = {Fan, Yi-Zhong and Han, Ming-Zhe and Jiang, Jin-Liang and Shao, Dong-Sheng and Tang, Shao-Peng},
	journal = {Phys. Rev. D},
	volume = {109},
	issue = {4},
	pages = {043052},
	numpages = {16},
	year = {2024},
	month = {Feb},
	publisher = {American Physical Society},
	doi = {10.1103/PhysRevD.109.043052},
	url = {https://link.aps.org/doi/10.1103/PhysRevD.109.043052}
}

@article{biswas-25a,
	title = {Simultaneously constraining the neutron star equation of state and mass distribution through multimessenger observations and nuclear benchmarks},
	author = {Biswas, Bhaskar and Rosswog, Stephan},
	journal = {Phys. Rev. D},
	volume = {112},
	issue = {2},
	pages = {023045},
	numpages = {23},
	year = {2025},
	month = {Jul},
	publisher = {American Physical Society},
	doi = {10.1103/8lv3-1ywb},
	url = {https://link.aps.org/doi/10.1103/8lv3-1ywb}
}

@article{golomb-25,
	title={Interplay of astrophysics and nuclear physics in determining the properties of neutron stars},
	volume={111},
	ISSN={2470-0029},
	url={http://dx.doi.org/10.1103/PhysRevD.111.023029},
	DOI={10.1103/physrevd.111.023029},
	number={2},
	journal={Physical Review D},
	publisher={American Physical Society (APS)},
	author={Golomb, Jacob and Legred, Isaac and Chatziioannou, Katerina and Landry, Philippe},
	year={2025},
	month=jan
}

@article{mamun-21,
    title = {Combining Electromagnetic and Gravitational-Wave Constraints on Neutron-Star Masses and Radii},
    author = {Al-Mamun, Mohammad and Steiner, Andrew W. and N\"attil\"a, Joonas and Lange, Jacob and O'Shaughnessy, Richard and Tews, Ingo and Gandolfi, Stefano and Heinke, Craig and Han, Sophia},
    journal = {Phys. Rev. Lett.},
    volume = {126},
    issue = {6},
    pages = {061101},
    numpages = {6},
    year = {2021},
    month = {Feb},
    publisher = {American Physical Society},
    doi = {10.1103/PhysRevLett.126.061101},
    url = {https://link.aps.org/doi/10.1103/PhysRevLett.126.061101}
}

@article{kiziltan-13,
    doi = {10.1088/0004-637x/778/1/66},
    url = {https://doi.org/10.1088/0004-637x/778/1/66},
    year = 2013,
    month = {nov},
    publisher = {American Astronomical Society},
    volume = {778},
    number = {1},
    pages = {66},
    author = {Bülent Kiziltan and Athanasios Kottas and Maria De Yoreo and Stephen E. Thorsett},
    title = {{The Neutron Star Mass Distribution}},
    journal = {The Astrophysical Journal},
}

@article{alsing-18,
    author = {{Alsing}, Justin and {Silva}, Hector O. and {Berti}, Emanuele},
    title = "{Evidence for a maximum mass cut-off in the neutron star mass distribution and constraints on the equation of state}",
    journal = {\mnras},
    year = 2018,
    month = jul,
    volume = {478},
    number = {1},
    pages = {1377-1391},
    doi = {10.1093/mnras/sty1065},
    archivePrefix = {arXiv},
    eprint = {1709.07889},
    primaryClass = {astro-ph.HE},
    adsurl = {https://ui.adsabs.harvard.edu/abs/2018MNRAS.478.1377A}
}

@article{cromartie-20,
    author = {{Cromartie}, H.~T. and {Fonseca}, E. and {Ransom}, S.~M. and {Demorest}, P.~B. and {Arzoumanian}, Z. and {Blumer}, H. and {Brook}, P.~R. and {DeCesar}, M.~E. and {Dolch}, T. and {Ellis}, J.~A. and {Ferdman}, R.~D. and {Ferrara}, E.~C. and {Garver-Daniels}, N. and {Gentile}, P.~A. and {Jones}, M.~L. and {Lam}, M.~T. and {Lorimer}, D.~R. and {Lynch}, R.~S. and {McLaughlin}, M.~A. and {Ng}, C. and {Nice}, D.~J. and {Pennucci}, T.~T. and {Spiewak}, R. and {Stairs}, I.~H. and {Stovall}, K. and {Swiggum}, J.~K. and {Zhu}, W.~W.},
    title = "{Relativistic Shapiro delay measurements of an extremely massive millisecond pulsar}",
    journal = {Nature Astronomy},
    year = 2020,
    month = jan,
    volume = {4},
    pages = {72-76},
    doi = {10.1038/s41550-019-0880-2},
    archivePrefix = {arXiv},
    eprint = {1904.06759},
    primaryClass = {astro-ph.HE},
    adsurl = {https://ui.adsabs.harvard.edu/abs/2020NatAs...4...72C}
}

@article{komoltsev-22,
    doi = {10.1103/physrevlett.128.202701},
    url = {https://doi.org/10.1103%2Fphysrevlett.128.202701},
    year = 2022,
    month = {may},
    publisher = {American Physical Society ({APS})},
    volume = {128},
    number = {20},
    author = {Oleg Komoltsev and Aleksi Kurkela},
    title = {How Perturbative {QCD} Constrains the Equation of State at Neutron-Star Densities}, 
    journal = {Physical Review Letters}
}

@article{ligo-17,
    title = {GW170817: Observation of Gravitational Waves from a Binary Neutron Star Inspiral},
    author = {Abbott, B. P. and others},
    collaboration = {LIGO Scientific Collaboration and Virgo Collaboration},
    journal = {Phys. Rev. Lett.},
    volume = {119},
    issue = {16},
    pages = {161101},
    numpages = {18},
    year = {2017},
    month = {Oct},
    publisher = {American Physical Society},
    doi = {10.1103/PhysRevLett.119.161101},
    url = {https://link.aps.org/doi/10.1103/PhysRevLett.119.161101}
}

@article{ligo-19,
    doi = {10.3847/2041-8213/ab75f5},
    url = {https://doi.org/10.3847%2F2041-8213%2Fab75f5},
    year = 2020,
    month = {mar},
    publisher = {American Astronomical Society},
    volume = {892},
    number = {1},
    pages = {L3},
    author = {Abbott, B. P. and others},
    collaboration = {LIGO Scientific Collaboration and Virgo Collaboration},
    title = {{GW}190425: Observation of a Compact Binary Coalescence with Total Mass~$\sim$~3.4 M$\less$sub$\greater$$\odot$$\less$/sub$\greater$},
    journal = {The Astrophysical Journal Letters}
}

@article{gandolfi-12,
    title = {Maximum mass and radius of neutron stars, and the nuclear symmetry energy},
    author = {Gandolfi, S. and Carlson, J. and Reddy, Sanjay},
    journal = {Phys. Rev. C},
    volume = {85},
    issue = {3},
    pages = {032801},
    numpages = {5},
    year = {2012},
    month = {Mar},
    publisher = {American Physical Society},
    doi = {10.1103/PhysRevC.85.032801},
    url = {https://link.aps.org/doi/10.1103/PhysRevC.85.032801}
}

@article{steiner-13,
	doi = {10.1088/2041-8205/765/1/L5},
	url = {https://doi.org/10.1088/2041-8205/765/1/L5},
	year = {2013},
	month = {feb},
	publisher = {The American Astronomical Society},
	volume = {765},
	number = {1},
	pages = {L5},
	author = {Steiner, Andrew W. and Lattimer, James M. and Brown, Edward F.},
	title = {THE NEUTRON STAR MASS–RADIUS RELATION AND THE EQUATION OF STATE OF DENSE MATTER},
	journal = {The Astrophysical Journal Letters}
}

@article{steiner-15,
  title = {Using neutron star observations to determine crust thicknesses, moments of inertia, and tidal deformabilities},
  author = {Steiner, A. W. and Gandolfi, S. and Fattoyev, F. J. and Newton, W. G.},
  journal = {Phys. Rev. C},
  volume = {91},
  issue = {1},
  pages = {015804},
  numpages = {7},
  year = {2015},
  month = {Jan},
  publisher = {American Physical Society},
  doi = {10.1103/PhysRevC.91.015804},
  url = {https://link.aps.org/doi/10.1103/PhysRevC.91.015804}
}

@article{steiner-16,
	title={Neutron star radii, universal relations, and the role of prior distributions},
	author={Steiner, Andrew W and Lattimer, James M and Brown, Edward F},
	journal={The European Physical Journal A},
	volume={52},
	number={2},
	pages={18},
	year={2016},
	publisher={Springer}
}

@article{nattila-16,
    author = {{N{\"a}ttil{\"a}}, J. and {Steiner}, A.~W. and {Kajava}, J.~J.~E. and {Suleimanov}, V.~F. and {Poutanen}, J.},
    title = "{Equation of state constraints for the cold dense matter inside neutron stars using the cooling tail method}",
    journal = {\aap},
    year = 2016,
    month = jun,
    volume = {591},
    eid = {A25},
    pages = {A25},
    doi = {10.1051/0004-6361/201527416},
    archivePrefix = {arXiv},
    eprint = {1509.06561},
    primaryClass = {astro-ph.HE},
    adsurl = {https://ui.adsabs.harvard.edu/abs/2016A&A...591A..25N}
}

@article{nattila-17,
    author = {{N{\"a}ttil{\"a}}, J. and {Miller}, M.~C. and {Steiner}, A.~W. and {Kajava}, J.~J.~E. and {Suleimanov}, V.~F. and {Poutanen}, J.},
    title = "{Neutron star mass and radius measurements from atmospheric model fits to X-ray burst cooling tail spectra}",
    journal = {\aap},
    year = 2017,
    month = dec,
    volume = {608},
    eid = {A31},
    pages = {A31},
    doi = {10.1051/0004-6361/201731082},
    archivePrefix = {arXiv},
    eprint = {1709.09120},
    primaryClass = {astro-ph.HE},
    adsurl = {https://ui.adsabs.harvard.edu/abs/2017A&A...608A..31N}
}

@article{steiner-18,
    author = {{Steiner}, A.~W. and {Heinke}, C.~O. and {Bogdanov}, S. and {Li}, C.~K. and {Ho}, W.~C.~G. and {Bahramian}, A. and {Han}, S.},
    title = "{Constraining the mass and radius of neutron stars in globular clusters}",
    journal = {\mnras},
    year = 2018,
    month = may,
    volume = {476},
    number = {1},
    pages = {421-435},
    doi = {10.1093/mnras/sty215},
    archivePrefix = {arXiv},
    eprint = {1709.05013},
    primaryClass = {astro-ph.HE},
    adsurl = {https://ui.adsabs.harvard.edu/abs/2018MNRAS.476..421S}
}

@article{riley-19,
    author = {{Riley}, T.~E. and {Watts}, A.~L. and {Bogdanov}, S. and {Ray}, P.~S. and {Ludlam}, R.~M. and {Guillot}, S. and {Arzoumanian}, Z. and {Baker}, C.~L. and {Bilous}, A.~V. and {Chakrabarty}, D. and {Gendreau}, K.~C. and {Harding}, A.~K. and {Ho}, W.~C.~G. and {Lattimer}, J.~M. and {Morsink}, S.~M. and {Strohmayer}, T.~E.},
    title = "{A NICER View of PSR J0030+0451: Millisecond Pulsar Parameter Estimation}",
    journal = {\apjl},
    year = 2019,
    month = dec,
    volume = {887},
    number = {1},
    eid = {L21},
    pages = {L21},
    doi = {10.3847/2041-8213/ab481c},
    archivePrefix = {arXiv},
    eprint = {1912.05702},
    primaryClass = {astro-ph.HE},
    adsurl = {https://ui.adsabs.harvard.edu/abs/2019ApJ...887L..21R}
}

@article{riley-21,
    author = {{Riley}, Thomas E. and {Watts}, Anna L. and {Ray}, Paul S. and {Bogdanov}, Slavko and {Guillot}, Sebastien and {Morsink}, Sharon M. and {Bilous}, Anna V. and {Arzoumanian}, Zaven and {Choudhury}, Devarshi and {Deneva}, Julia S. and {Gendreau}, Keith C. and {Harding}, Alice K. and {Ho}, Wynn C.~G. and {Lattimer}, James M. and {Loewenstein}, Michael and {Ludlam}, Renee M. and {Markwardt}, Craig B. and {Okajima}, Takashi and {Prescod-Weinstein}, Chanda and {Remillard}, Ronald A. and {Wolff}, Michael T. and {Fonseca}, Emmanuel and {Cromartie}, H. Thankful and {Kerr}, Matthew and {Pennucci}, Timothy T. and {Parthasarathy}, Aditya and {Ransom}, Scott and {Stairs}, Ingrid and {Guillemot}, Lucas and {Cognard}, Ismael},
    title = "{A NICER View of the Massive Pulsar PSR J0740+6620 Informed by Radio Timing and XMM-Newton Spectroscopy}",
    journal = {\apjl},
    year = 2021,
    month = sep,
    volume = {918},
    number = {2},
    eid = {L27},
    pages = {L27},
    doi = {10.3847/2041-8213/ac0a81},
    archivePrefix = {arXiv},
    eprint = {2105.06980},
    primaryClass = {astro-ph.HE},
    adsurl = {https://ui.adsabs.harvard.edu/abs/2021ApJ...918L..27R}
}

@article{miller-19,
    author = {{Miller}, M.~C. and {Lamb}, F.~K. and {Dittmann}, A.~J. and {Bogdanov}, S. and {Arzoumanian}, Z. and {Gendreau}, K.~C. and {Guillot}, S. and {Harding}, A.~K. and {Ho}, W.~C.~G. and {Lattimer}, J.~M. and {Ludlam}, R.~M. and {Mahmoodifar}, S. and {Morsink}, S.~M. and {Ray}, P.~S. and {Strohmayer}, T.~E. and {Wood}, K.~S. and {Enoto}, T. and {Foster}, R. and {Okajima}, T. and {Prigozhin}, G. and {Soong}, Y.},
    title = "{PSR J0030+0451 Mass and Radius from NICER Data and Implications for the Properties of Neutron Star Matter}",
    journal = {\apjl},
    year = 2019,
    month = dec,
    volume = {887},
    number = {1},
    eid = {L24},
    pages = {L24},
    doi = {10.3847/2041-8213/ab50c5},
    archivePrefix = {arXiv},
    eprint = {1912.05705},
    primaryClass = {astro-ph.HE},
    adsurl = {https://ui.adsabs.harvard.edu/abs/2019ApJ...887L..24M}
}

@article{miller-21,
    author = {{Miller}, M.~C. and {Lamb}, F.~K. and {Dittmann}, A.~J. and {Bogdanov}, S. and {Arzoumanian}, Z. and {Gendreau}, K.~C. and {Guillot}, S. and {Ho}, W.~C.~G. and {Lattimer}, J.~M. and {Loewenstein}, M. and {Morsink}, S.~M. and {Ray}, P.~S. and {Wolff}, M.~T. and {Baker}, C.~L. and {Cazeau}, T. and {Manthripragada}, S. and {Markwardt}, C.~B. and {Okajima}, T. and {Pollard}, S. and {Cognard}, I. and {Cromartie}, H.~T. and {Fonseca}, E. and {Guillemot}, L. and {Kerr}, M. and {Parthasarathy}, A. and {Pennucci}, T.~T. and {Ransom}, S. and {Stairs}, I.},
    title = "{The Radius of PSR J0740+6620 from NICER and XMM-Newton Data}",
    journal = {\apjl},
    year = 2021,
    month = sep,
    volume = {918},
    number = {2},
    eid = {L28},
    pages = {L28},
    doi = {10.3847/2041-8213/ac089b},
    archivePrefix = {arXiv},
    eprint = {2105.06979},
    primaryClass = {astro-ph.HE},
    adsurl = {https://ui.adsabs.harvard.edu/abs/2021ApJ...918L..28M}
}

@article{farrow-19,
    doi = {10.3847/1538-4357/ab12e3},
    url = {https://dx.doi.org/10.3847/1538-4357/ab12e3},
    year = {2019},
    month = {apr},
    publisher = {The American Astronomical Society},
    volume = {876},
    number = {1},
    pages = {18},
    author = {Nicholas Farrow and Xing-Jiang Zhu and Eric Thrane},
    title = {The Mass Distribution of Galactic Double Neutron Stars},
    journal = {The Astrophysical Journal}
}

@article{shao-20,
    title = {Maximum mass cutoff in the neutron star mass distribution and the prospect of forming supramassive objects in the double neutron star mergers},
    author = {Shao, Dong-Sheng and Tang, Shao-Peng and Jiang, Jin-Liang and Fan, Yi-Zhong},
    journal = {Phys. Rev. D},
    volume = {102},
    issue = {6},
    pages = {063006},
    numpages = {16},
    year = {2020},
    month = {Sep},
    publisher = {American Physical Society},
    doi = {10.1103/PhysRevD.102.063006},
    url = {https://link.aps.org/doi/10.1103/PhysRevD.102.063006}
}

@article{valentim-11,
    author = {Valentim, R. and Rangel, E. and Horvath, J. E.},
    title = "{On the mass distribution of neutron stars}",
    journal = {Monthly Notices of the Royal Astronomical Society},
    volume = {414},
    number = {2},
    pages = {1427-1431},
    year = {2011},
    month = {06},
    issn = {0035-8711},
    doi = {10.1111/j.1365-2966.2011.18477.x},
    url = {https://doi.org/10.1111/j.1365-2966.2011.18477.x}
}

@article{ozel-12,
    doi = {10.1088/0004-637X/757/1/55},
    url = {https://dx.doi.org/10.1088/0004-637X/757/1/55},
    year = {2012},
    month = {sep},
    publisher = {The American Astronomical Society},
    volume = {757},
    number = {1},
    pages = {55},
    author = {Feryal \"Ozel and Dimitrios Psaltis and Ramesh Narayan and Antonio Santos Villarreal},
    title = {On the Mass Distribution and Birth Masses of Neutron Stars},
    journal = {The Astrophysical Journal}
}

@article{chatziioannou-20,
    title = {Inferring the maximum and minimum mass of merging neutron stars with gravitational waves},
    author = {Chatziioannou, Katerina and Farr, Will M.},
    journal = {Phys. Rev. D},
    volume = {102},
    issue = {6},
    pages = {064063},
    numpages = {8},
    year = {2020},
    month = {Sep},
    publisher = {American Physical Society},
    doi = {10.1103/PhysRevD.102.064063},
    url = {https://link.aps.org/doi/10.1103/PhysRevD.102.064063}
}

@article{keller-23,
    title = {Nuclear Equation of State for Arbitrary Proton Fraction and Temperature Based on Chiral Effective Field Theory and a Gaussian Process Emulator},
    author = {Keller, J. and Hebeler, K. and Schwenk, A.},
    journal = {Phys. Rev. Lett.},
    volume = {130},
    issue = {7},
    pages = {072701},
    numpages = {6},
    year = {2023},
    month = {Feb},
    publisher = {American Physical Society},
    doi = {10.1103/PhysRevLett.130.072701},
    url = {https://link.aps.org/doi/10.1103/PhysRevLett.130.072701}
}

@article{martinez-15,
    author = {{Martinez}, J.~G. and {Stovall}, K. and {Freire}, P.~C.~C. and {Deneva}, J.~S. and {Jenet}, F.~A. and {McLaughlin}, M.~A. and {Bagchi}, M. and {Bates}, S.~D. and {Ridolfi}, A.},
    title = "{Pulsar J0453+1559: A Double Neutron Star System with a Large Mass Asymmetry}",
    journal = {\apj},
    year = 2015,
    month = oct,
    volume = {812},
    number = {2},
    eid = {143},
    pages = {143},
    doi = {10.1088/0004-637X/812/2/143},
    archivePrefix = {arXiv},
    eprint = {1509.08805},
    primaryClass = {astro-ph.HE},
    adsurl = {https://ui.adsabs.harvard.edu/abs/2015ApJ...812..143M}
}

@article{vanleeuwen-15,
    author = {{van Leeuwen}, J. and {Kasian}, L. and {Stairs}, I.~H. and {Lorimer}, D.~R. and {Camilo}, F. and {Chatterjee}, S. and {Cognard}, I. and {Desvignes}, G. and {Freire}, P.~C.~C. and {Janssen}, G.~H. and {Kramer}, M. and {Lyne}, A.~G. and {Nice}, D.~J. and {Ransom}, S.~M. and {Stappers}, B.~W. and {Weisberg}, J.~M.},
    title = "{The Binary Companion of Young, Relativistic Pulsar J1906+0746}",
    journal = {\apj},
    year = 2015,
    month = jan,
    volume = {798},
    number = {2},
    eid = {118},
    pages = {118},
    doi = {10.1088/0004-637X/798/2/118},
    archivePrefix = {arXiv},
    eprint = {1411.1518},
    primaryClass = {astro-ph.SR},
    adsurl = {https://ui.adsabs.harvard.edu/abs/2015ApJ...798..118V}
}

@article{fonseca-14,
    author = {{Fonseca}, Emmanuel and {Stairs}, Ingrid H. and {Thorsett}, Stephen E.},
    title = "{A Comprehensive Study of Relativistic Gravity Using PSR B1534+12}",
    journal = {\apj},
    year = 2014,
    month = may,
    volume = {787},
    number = {1},
    eid = {82},
    pages = {82},
    doi = {10.1088/0004-637X/787/1/82},
    archivePrefix = {arXiv},
    eprint = {1402.4836},
    primaryClass = {astro-ph.HE},
    adsurl = {https://ui.adsabs.harvard.edu/abs/2014ApJ...787...82F}
}

@article{weisberg-10,
    author = {{Weisberg}, J.~M. and {Nice}, D.~J. and {Taylor}, J.~H.},
    title = "{Timing Measurements of the Relativistic Binary Pulsar PSR B1913+16}",
    journal = {\apj},
    year = 2010,
    month = oct,
    volume = {722},
    number = {2},
    pages = {1030-1034},
    doi = {10.1088/0004-637X/722/2/1030},
    archivePrefix = {arXiv},
    eprint = {1011.0718},
    primaryClass = {astro-ph.GA},
    adsurl = {https://ui.adsabs.harvard.edu/abs/2010ApJ...722.1030W}
}

@article{jacoby-06,
    author = {{Jacoby}, B.~A. and {Cameron}, P.~B. and {Jenet}, F.~A. and {Anderson}, S.~B. and {Murty}, R.~N. and {Kulkarni}, S.~R.},
    title = "{Measurement of Orbital Decay in the Double Neutron Star Binary PSR B2127+11C}",
    journal = {\apjl},
    year = 2006,
    month = jun,
    volume = {644},
    number = {2},
    pages = {L113-L116},
    doi = {10.1086/505742},
    archivePrefix = {arXiv},
    eprint = {astro-ph/0605375},
    primaryClass = {astro-ph},
    adsurl = {https://ui.adsabs.harvard.edu/abs/2006ApJ...644L.113J}
}

@article{kramer-06,
    author = {{Kramer}, M. and {Stairs}, I.~H. and {Manchester}, R.~N. and {McLaughlin}, M.~A. and {Lyne}, A.~G. and {Ferdman}, R.~D. and {Burgay}, M. and {Lorimer}, D.~R. and {Possenti}, A. and {D'Amico}, N. and {Sarkissian}, J.~M. and {Hobbs}, G.~B. and {Reynolds}, J.~E. and {Freire}, P.~C.~C. and {Camilo}, F.},
    title = "{Tests of General Relativity from Timing the Double Pulsar}",
    journal = {Science},
    year = 2006,
    month = oct,
    volume = {314},
    number = {5796},
    pages = {97-102},
    doi = {10.1126/science.1132305},
    archivePrefix = {arXiv},
    eprint = {astro-ph/0609417},
    primaryClass = {astro-ph},
    adsurl = {https://ui.adsabs.harvard.edu/abs/2006Sci...314...97K}
}

@article{ferdman-14,
    author = {{Ferdman}, R.~D. and {Stairs}, I.~H. and {Kramer}, M. and {Janssen}, G.~H. and {Bassa}, C.~G. and {Stappers}, B.~W. and {Demorest}, P.~B. and {Cognard}, I. and {Desvignes}, G. and {Theureau}, G. and {Burgay}, M. and {Lyne}, A.~G. and {Manchester}, R.~N. and {Possenti}, A.},
    title = "{PSR J1756-2251: a pulsar with a low-mass neutron star companion}",
    journal = {\mnras},
    year = 2014,
    month = sep,
    volume = {443},
    number = {3},
    pages = {2183-2196},
    doi = {10.1093/mnras/stu1223},
    archivePrefix = {arXiv},
    eprint = {1406.5507},
    primaryClass = {astro-ph.SR},
    adsurl = {https://ui.adsabs.harvard.edu/abs/2014MNRAS.443.2183F}
}

@article{lynch-12,
    author = {{Lynch}, Ryan S. and {Freire}, Paulo C.~C. and {Ransom}, Scott M. and {Jacoby}, Bryan A.},
    title = "{The Timing of Nine Globular Cluster Pulsars}",
    journal = {\apj},
    year = 2012,
    month = feb,
    volume = {745},
    number = {2},
    eid = {109},
    pages = {109},
    doi = {10.1088/0004-637X/745/2/109},
    archivePrefix = {arXiv},
    eprint = {1112.2612},
    primaryClass = {astro-ph.HE},
    adsurl = {https://ui.adsabs.harvard.edu/abs/2012ApJ...745..109L}
}

@inproceedings{nice-08,
    author = {{Nice}, David J. and {Stairs}, Ingrid H. and {Kasian}, Laura E.},
    title = "{Masses of Neutron Stars in Binary Pulsar Systems}",
    booktitle = {40 Years of Pulsars: Millisecond Pulsars, Magnetars and More},
    year = 2008,
    editor = {{Bassa}, C. and {Wang}, Z. and {Cumming}, A. and {Kaspi}, V.~M.},
    series = {American Institute of Physics Conference Series},
    volume = {983},
    month = feb,
    pages = {453-458},
    doi = {10.1063/1.2900273},
    adsurl = {https://ui.adsabs.harvard.edu/abs/2008AIPC..983..453N}
}

@article{ferdman-10,
    author = {{Ferdman}, R.~D. and {Stairs}, I.~H. and {Kramer}, M. and {McLaughlin}, M.~A. and {Lorimer}, D.~R. and {Nice}, D.~J. and {Manchester}, R.~N. and {Hobbs}, G. and {Lyne}, A.~G. and {Camilo}, F. and {Possenti}, A. and {Demorest}, P.~B. and {Cognard}, I. and {Desvignes}, G. and {Theureau}, G. and {Faulkner}, A. and {Backer}, D.~C.},
    title = "{A Precise Mass Measurement of the Intermediate-Mass Binary Pulsar PSR J1802 - 2124}",
    journal = {\apj},
    year = 2010,
    month = mar,
    volume = {711},
    number = {2},
    pages = {764-771},
    doi = {10.1088/0004-637X/711/2/764},
    archivePrefix = {arXiv},
    eprint = {1002.0514},
    primaryClass = {astro-ph.SR},
    adsurl = {https://ui.adsabs.harvard.edu/abs/2010ApJ...711..764F}
}

@article{stairs-06,
    author = {{Stairs}, I.~H.},
    title = "{Masses of radio pulsars}",
    journal = {Journal of Physics G Nuclear Physics},
    year = 2006,
    month = dec,
    volume = {32},
    number = {12},
    pages = {S259-S265},
    doi = {10.1088/0954-3899/32/12/S32},
    adsurl = {https://ui.adsabs.harvard.edu/abs/2006JPhG...32S.259S}
}

@article{corongiu-07,
    author = {{Corongiu}, A. and {Kramer}, M. and {Stappers}, B.~W. and {Lyne}, A.~G. and {Jessner}, A. and {Possenti}, A. and {D'Amico}, N. and {L{\"o}hmer}, O.},
    title = "{The binary pulsar PSR J1811-1736: evidence of a low amplitude supernova kick}",
    journal = {\aap},
    year = 2007,
    month = feb,
    volume = {462},
    number = {2},
    pages = {703-709},
    doi = {10.1051/0004-6361:20054385},
    archivePrefix = {arXiv},
    eprint = {astro-ph/0611436},
    primaryClass = {astro-ph},
    adsurl = {https://ui.adsabs.harvard.edu/abs/2007A&A...462..703C}
}

@article{champion-05,
    author = {{Champion}, D.~J. and {Lorimer}, D.~R. and {McLaughlin}, M.~A. and {Xilouris}, K.~M. and {Arzoumanian}, Z. and {Freire}, P.~C.~C. and {Lommen}, A.~N. and {Cordes}, J.~M. and {Camilo}, F.},
    title = "{Arecibo timing and single-pulse observations of 17 pulsars}",
    journal = {\mnras},
    year = 2005,
    month = nov,
    volume = {363},
    number = {3},
    pages = {929-936},
    doi = {10.1111/j.1365-2966.2005.09499.x},
    archivePrefix = {arXiv},
    eprint = {astro-ph/0508320},
    primaryClass = {astro-ph},
    adsurl = {https://ui.adsabs.harvard.edu/abs/2005MNRAS.363..929C}
}

@article{berezina-17,
    author = {{Berezina}, M. and {Champion}, D.~J. and {Freire}, P.~C.~C. and {Tauris}, T.~M. and {Kramer}, M. and {Lyne}, A.~G. and {Stappers}, B.~W. and {Guillemot}, L. and {Cognard}, I. and {Barr}, E.~D. and {Eatough}, R.~P. and {Karuppusamy}, R. and {Spitler}, L.~G. and {Desvignes}, G.},
    title = "{The discovery of two mildly recycled binary pulsars in the Northern High Time Resolution Universe pulsar survey}",
    journal = {\mnras},
    year = 2017,
    month = oct,
    volume = {470},
    number = {4},
    pages = {4421-4433},
    doi = {10.1093/mnras/stx1518},
    archivePrefix = {arXiv},
    eprint = {1706.06417},
    primaryClass = {astro-ph.HE},
    adsurl = {https://ui.adsabs.harvard.edu/abs/2017MNRAS.470.4421B}
}

@article{arzoumanian-18,
    author = {{Arzoumanian}, Z. and {Baker}, P.~T. and {Brazier}, A. and {Burke-Spolaor}, S. and {Chamberlin}, S.~J. and {Chatterjee}, S. and {Christy}, B. and {Cordes}, J.~M. and {Cornish}, N.~J. and {Crawford}, F. and {Thankful Cromartie}, H. and {Crowter}, K. and {DeCesar}, M. and {Demorest}, P.~B. and {Dolch}, T. and {Ellis}, J.~A. and {Ferdman}, R.~D. and {Ferrara}, E. and {Folkner}, W.~M. and {Fonseca}, E. and {Garver-Daniels}, N. and {Gentile}, P.~A. and {Haas}, R. and {Hazboun}, J.~S. and {Huerta}, E.~A. and {Islo}, K. and {Jones}, G. and {Jones}, M.~L. and {Kaplan}, D.~L. and {Kaspi}, V.~M. and {Lam}, M.~T. and {Lazio}, T.~J.~W. and {Levin}, L. and {Lommen}, A.~N. and {Lorimer}, D.~R. and {Luo}, J. and {Lynch}, R.~S. and {Madison}, D.~R. and {McLaughlin}, M.~A. and {McWilliams}, S.~T. and {Mingarelli}, C.~M.~F. and {Ng}, C. and {Nice}, D.~J. and {Park}, R.~S. and {Pennucci}, T.~T. and {Pol}, N.~S. and {Ransom}, S.~M. and {Ray}, P.~S. and {Rasskazov}, A. and {Siemens}, X. and {Simon}, J. and {Spiewak}, R. and {Stairs}, I.~H. and {Stinebring}, D.~R. and {Stovall}, K. and {Swiggum}, J. and {Taylor}, S.~R. and {Vallisneri}, M. and {van Haasteren}, R. and {Vigeland}, S. and {Zhu}, W.~W. and {NANOGrav Collaboration}},
    title = "{The NANOGrav 11 Year Data Set: Pulsar-timing Constraints on the Stochastic Gravitational-wave Background}",
    journal = {\apj},
    year = 2018,
    month = may,
    volume = {859},
    number = {1},
    eid = {47},
    pages = {47},
    doi = {10.3847/1538-4357/aabd3b},
    archivePrefix = {arXiv},
    eprint = {1801.02617},
    primaryClass = {astro-ph.HE},
    adsurl = {https://ui.adsabs.harvard.edu/abs/2018ApJ...859...47A}
}

@article{desvignes-16,
    author = {{Desvignes}, G. and {Caballero}, R.~N. and {Lentati}, L. and {Verbiest}, J.~P.~W. and {Champion}, D.~J. and {Stappers}, B.~W. and {Janssen}, G.~H. and {Lazarus}, P. and {Os{\l}owski}, S. and {Babak}, S. and {Bassa}, C.~G. and {Brem}, P. and {Burgay}, M. and {Cognard}, I. and {Gair}, J.~R. and {Graikou}, E. and {Guillemot}, L. and {Hessels}, J.~W.~T. and {Jessner}, A. and {Jordan}, C. and {Karuppusamy}, R. and {Kramer}, M. and {Lassus}, A. and {Lazaridis}, K. and {Lee}, K.~J. and {Liu}, K. and {Lyne}, A.~G. and {McKee}, J. and {Mingarelli}, C.~M.~F. and {Perrodin}, D. and {Petiteau}, A. and {Possenti}, A. and {Purver}, M.~B. and {Rosado}, P.~A. and {Sanidas}, S. and {Sesana}, A. and {Shaifullah}, G. and {Smits}, R. and {Taylor}, S.~R. and {Theureau}, G. and {Tiburzi}, C. and {van Haasteren}, R. and {Vecchio}, A.},
    title = "{High-precision timing of 42 millisecond pulsars with the European Pulsar Timing Array}",
    journal = {\mnras},
    year = 2016,
    month = may,
    volume = {458},
    number = {3},
    pages = {3341-3380},
    doi = {10.1093/mnras/stw483},
    archivePrefix = {arXiv},
    eprint = {1602.08511},
    primaryClass = {astro-ph.HE},
    adsurl = {https://ui.adsabs.harvard.edu/abs/2016MNRAS.458.3341D}
}

@article{bhat-08,
    author = {{Bhat}, N.~D. Ramesh and {Bailes}, Matthew and {Verbiest}, Joris P.~W.},
    title = "{Gravitational-radiation losses from the pulsar white-dwarf binary PSR J1141 6545}",
    journal = {\prd},
    year = 2008,
    month = jun,
    volume = {77},
    number = {12},
    eid = {124017},
    pages = {124017},
    doi = {10.1103/PhysRevD.77.124017},
    archivePrefix = {arXiv},
    eprint = {0804.0956},
    primaryClass = {astro-ph},
    adsurl = {https://ui.adsabs.harvard.edu/abs/2008PhRvD..77l4017B}
}

@article{antoniadis-12,
    author = {{Antoniadis}, J. and {van Kerkwijk}, M.~H. and {Koester}, D. and {Freire}, P.~C.~C. and {Wex}, N. and {Tauris}, T.~M. and {Kramer}, M. and {Bassa}, C.~G.},
    title = "{The relativistic pulsar-white dwarf binary PSR J1738+0333 - I. Mass determination and evolutionary history}",
    journal = {\mnras},
    year = 2012,
    month = jul,
    volume = {423},
    number = {4},
    pages = {3316-3327},
    doi = {10.1111/j.1365-2966.2012.21124.x},
    archivePrefix = {arXiv},
    eprint = {1204.3948},
    primaryClass = {astro-ph.HE},
    adsurl = {https://ui.adsabs.harvard.edu/abs/2012MNRAS.423.3316A}
}

@article{antoniadis-13,
    author = {{Antoniadis}, John and {Freire}, Paulo C.~C. and {Wex}, Norbert and {Tauris}, Thomas M. and {Lynch}, Ryan S. and {van Kerkwijk}, Marten H. and {Kramer}, Michael and {Bassa}, Cees and {Dhillon}, Vik S. and {Driebe}, Thomas and {Hessels}, Jason W.~T. and {Kaspi}, Victoria M. and {Kondratiev}, Vladislav I. and {Langer}, Norbert and {Marsh}, Thomas R. and {McLaughlin}, Maura A. and {Pennucci}, Timothy T. and {Ransom}, Scott M. and {Stairs}, Ingrid H. and {van Leeuwen}, Joeri and {Verbiest}, Joris P.~W. and {Whelan}, David G.},
    title = "{A Massive Pulsar in a Compact Relativistic Binary}",
    journal = {Science},
    year = 2013,
    month = apr,
    volume = {340},
    number = {6131},
    pages = {448},
    doi = {10.1126/science.1233232},
    archivePrefix = {arXiv},
    eprint = {1304.6875},
    primaryClass = {astro-ph.HE},
    adsurl = {https://ui.adsabs.harvard.edu/abs/2013Sci...340..448A}
}

@article{cognard-17,
    author = {{Cognard}, Isma{\"e}l and {Freire}, Paulo C.~C. and {Guillemot}, Lucas and {Theureau}, Gilles and {Tauris}, Thomas M. and {Wex}, Norbert and {Graikou}, Eleni and {Kramer}, Michael and {Stappers}, Benjamin and {Lyne}, Andrew G. and {Bassa}, Cees and {Desvignes}, Gregory and {Lazarus}, Patrick},
    title = "{A Massive-born Neutron Star with a Massive White Dwarf Companion}",
    journal = {\apj},
    year = 2017,
    month = aug,
    volume = {844},
    number = {2},
    eid = {128},
    pages = {128},
    doi = {10.3847/1538-4357/aa7bee},
    archivePrefix = {arXiv},
    eprint = {1706.08060},
    primaryClass = {astro-ph.HE},
    adsurl = {https://ui.adsabs.harvard.edu/abs/2017ApJ...844..128C}
}

@article{stovall-19,
    author = {{Stovall}, K. and {Freire}, P.~C.~C. and {Antoniadis}, J. and {Bagchi}, M. and {Deneva}, J.~S. and {Garver-Daniels}, N. and {Martinez}, J.~G. and {McLaughlin}, M.~A. and {Arzoumanian}, Z. and {Blumer}, H. and {Brook}, P.~R. and {Cromartie}, H.~T. and {Demorest}, P.~B. and {DeCesar}, M.~E. and {Dolch}, T. and {Ellis}, J.~A. and {Ferdman}, R.~D. and {Ferrara}, E.~C. and {Fonseca}, E. and {Gentile}, P.~A. and {Jones}, M.~L. and {Lam}, M.~T. and {Lorimer}, D.~R. and {Lynch}, R.~S. and {Ng}, C. and {Nice}, D.~J. and {Pennucci}, T.~T. and {Ransom}, S.~M. and {Spiewak}, R. and {Stairs}, I.~H. and {Swiggum}, J.~K. and {Vigeland}, S.~J. and {Zhu}, W.~W.},
    title = "{PSR J2234+0611: A New Laboratory for Stellar Evolution}",
    journal = {\apj},
    year = 2019,
    month = jan,
    volume = {870},
    number = {2},
    eid = {74},
    pages = {74},
    doi = {10.3847/1538-4357/aaf37d},
    archivePrefix = {arXiv},
    eprint = {1809.05064},
    primaryClass = {astro-ph.HE},
    adsurl = {https://ui.adsabs.harvard.edu/abs/2019ApJ...870...74S}
}

@article{deneva-12,
    author = {{Deneva}, J.~S. and {Freire}, P.~C.~C. and {Cordes}, J.~M. and {Lyne}, A.~G. and {Ransom}, S.~M. and {Cognard}, I. and {Camilo}, F. and {Nice}, D.~J. and {Stairs}, I.~H. and {Allen}, B. and {Bhat}, N.~D.~R. and {Bogdanov}, S. and {Brazier}, A. and {Champion}, D.~J. and {Chatterjee}, S. and {Crawford}, F. and {Desvignes}, G. and {Hessels}, J.~W.~T. and {Jenet}, F.~A. and {Kaspi}, V.~M. and {Knispel}, B. and {Kramer}, M. and {Lazarus}, P. and {van Leeuwen}, J. and {Lorimer}, D.~R. and {Lynch}, R.~S. and {McLaughlin}, M.~A. and {Scholz}, P. and {Siemens}, X. and {Stappers}, B.~W. and {Stovall}, K. and {Venkataraman}, A.},
    title = "{Two Millisecond Pulsars Discovered by the PALFA Survey and a Shapiro Delay Measurement}",
    journal = {\apj},
    year = 2012,
    month = sep,
    volume = {757},
    number = {1},
    eid = {89},
    pages = {89},
    doi = {10.1088/0004-637X/757/1/89},
    archivePrefix = {arXiv},
    eprint = {1208.1228},
    primaryClass = {astro-ph.SR},
    adsurl = {https://ui.adsabs.harvard.edu/abs/2012ApJ...757...89D}
}

@article{antoniadis-16,
    author = {{Antoniadis}, John and {Tauris}, Thomas M. and {Ozel}, Feryal and {Barr}, Ewan and {Champion}, David J. and {Freire}, Paulo C.~C.},
    title = "{The millisecond pulsar mass distribution: Evidence for bimodality and constraints on the maximum neutron star mass}",
    journal = {arXiv e-prints},
    year = 2016,
    month = may,
    eid = {arXiv:1605.01665},
    pages = {arXiv:1605.01665},
    doi = {10.48550/arXiv.1605.01665},
    archivePrefix = {arXiv},
    eprint = {1605.01665},
    primaryClass = {astro-ph.HE},
    adsurl = {https://ui.adsabs.harvard.edu/abs/2016arXiv160501665A}
}

@article{reardon-16,
    author = {{Reardon}, D.~J. and {Hobbs}, G. and {Coles}, W. and {Levin}, Y. and {Keith}, M.~J. and {Bailes}, M. and {Bhat}, N.~D.~R. and {Burke-Spolaor}, S. and {Dai}, S. and {Kerr}, M. and {Lasky}, P.~D. and {Manchester}, R.~N. and {Os{\l}owski}, S. and {Ravi}, V. and {Shannon}, R.~M. and {van Straten}, W. and {Toomey}, L. and {Wang}, J. and {Wen}, L. and {You}, X.~P. and {Zhu}, X. -J.},
    title = "{Timing analysis for 20 millisecond pulsars in the Parkes Pulsar Timing Array}",
    journal = {\mnras},
    year = 2016,
    month = jan,
    volume = {455},
    number = {2},
    pages = {1751-1769},
    doi = {10.1093/mnras/stv2395},
    archivePrefix = {arXiv},
    eprint = {1510.04434},
    primaryClass = {astro-ph.HE},
    adsurl = {https://ui.adsabs.harvard.edu/abs/2016MNRAS.455.1751R}
}

@article{bassa-06,
    author = {{Bassa}, C.~G. and {van Kerkwijk}, M.~H. and {Koester}, D. and {Verbunt}, F.},
    title = "{The masses of PSR J1911-5958A and its white dwarf companion}",
    journal = {\aap},
    year = 2006,
    month = sep,
    volume = {456},
    number = {1},
    pages = {295-304},
    doi = {10.1051/0004-6361:20065181},
    archivePrefix = {arXiv},
    eprint = {astro-ph/0603267},
    primaryClass = {astro-ph},
    adsurl = {https://ui.adsabs.harvard.edu/abs/2006A&A...456..295B}
}

@article{ransom-14,
    author = {{Ransom}, S.~M. and {Stairs}, I.~H. and {Archibald}, A.~M. and {Hessels}, J.~W.~T. and {Kaplan}, D.~L. and {van Kerkwijk}, M.~H. and {Boyles}, J. and {Deller}, A.~T. and {Chatterjee}, S. and {Schechtman-Rook}, A. and {Berndsen}, A. and {Lynch}, R.~S. and {Lorimer}, D.~R. and {Karako-Argaman}, C. and {Kaspi}, V.~M. and {Kondratiev}, V.~I. and {McLaughlin}, M.~A. and {van Leeuwen}, J. and {Rosen}, R. and {Roberts}, M.~S.~E. and {Stovall}, K.},
    title = "{A millisecond pulsar in a stellar triple system}",
    journal = {\nat},
    year = 2014,
    month = jan,
    volume = {505},
    number = {7484},
    pages = {520-524},
    doi = {10.1038/nature12917},
    archivePrefix = {arXiv},
    eprint = {1401.0535},
    primaryClass = {astro-ph.SR},
    adsurl = {https://ui.adsabs.harvard.edu/abs/2014Natur.505..520R}
}

@article{barr-17,
    author = {{Barr}, E.~D. and {Freire}, P.~C.~C. and {Kramer}, M. and {Champion}, D.~J. and {Berezina}, M. and {Bassa}, C.~G. and {Lyne}, A.~G. and {Stappers}, B.~W.},
    title = "{A massive millisecond pulsar in an eccentric binary}",
    journal = {\mnras},
    year = 2017,
    month = feb,
    volume = {465},
    number = {2},
    pages = {1711-1719},
    doi = {10.1093/mnras/stw2947},
    archivePrefix = {arXiv},
    eprint = {1611.03658},
    primaryClass = {astro-ph.HE},
    adsurl = {https://ui.adsabs.harvard.edu/abs/2017MNRAS.465.1711B}
}

@article{casares-10,
    author = {{Casares}, J. and {Gonz{\'a}lez Hern{\'a}ndez}, J.~I. and {Israelian}, G. and {Rebolo}, R.},
    title = "{On the mass of the neutron star in Cyg X-2}",
    journal = {\mnras},
    year = 2010,
    month = feb,
    volume = {401},
    number = {4},
    pages = {2517-2520},
    doi = {10.1111/j.1365-2966.2009.15828.x},
    archivePrefix = {arXiv},
    eprint = {0910.4496},
    primaryClass = {astro-ph.GA},
    adsurl = {https://ui.adsabs.harvard.edu/abs/2010MNRAS.401.2517C}
}

@inproceedings{gelino-02,
    author = {{Gelino}, D.~M. and {Tomsick}, J.~A. and {Heindl}, W.~A.},
    title = "{Measuring the Orbital Inclination Angle for the Low-Mass X-Ray Binary XTE J2123-058}",
    booktitle = {American Astronomical Society Meeting Abstracts},
    year = 2002,
    series = {American Astronomical Society Meeting Abstracts},
    volume = {201},
    month = dec,
    eid = {54.05},
    pages = {54.05},
    adsurl = {https://ui.adsabs.harvard.edu/abs/2002AAS...201.5405G}
}

@article{munozdarias-05,
    author = {{Mu{\~n}oz-Darias}, T. and {Casares}, J. and {Mart{\'\i}nez-Pais}, I.~G.},
    title = "{The ``K-Correction'' for Irradiated Emission Lines in LMXBs: Evidence for a Massive Neutron Star in X1822-371 (V691 CrA)}",
    journal = {\apj},
    year = 2005,
    month = dec,
    volume = {635},
    number = {1},
    pages = {502-507},
    doi = {10.1086/497420},
    archivePrefix = {arXiv},
    eprint = {astro-ph/0508547},
    primaryClass = {astro-ph},
    adsurl = {https://ui.adsabs.harvard.edu/abs/2005ApJ...635..502M}
}

@article{rawls-11,
    author = {{Rawls}, Meredith L. and {Orosz}, Jerome A. and {McClintock}, Jeffrey E. and {Torres}, Manuel A.~P. and {Bailyn}, Charles D. and {Buxton}, Michelle M.},
    title = "{Refined Neutron Star Mass Determinations for Six Eclipsing X-Ray Pulsar Binaries}",
    journal = {\apj},
    year = 2011,
    month = mar,
    volume = {730},
    number = {1},
    eid = {25},
    pages = {25},
    doi = {10.1088/0004-637X/730/1/25},
    archivePrefix = {arXiv},
    eprint = {1101.2465},
    primaryClass = {astro-ph.SR},
    adsurl = {https://ui.adsabs.harvard.edu/abs/2011ApJ...730...25R}
}

@article{steeghs-07,
    author = {{Steeghs}, D. and {Jonker}, P.~G.},
    title = "{On the Mass of the Neutron Star in V395 Carinae/2S 0921-630}",
    journal = {\apjl},
    year = 2007,
    month = nov,
    volume = {669},
    number = {2},
    pages = {L85-L88},
    doi = {10.1086/523848},
    archivePrefix = {arXiv},
    eprint = {0707.2067},
    primaryClass = {astro-ph},
    adsurl = {https://ui.adsabs.harvard.edu/abs/2007ApJ...669L..85S}
}
\bibliographystyle{aasjournalv7}

\appendix
\renewcommand{\theHequation}{\Alph{section}.\arabic{equation}}

\section{Skewed and Asymmetric Normal Distributions}
\label{app:a}
The probability distribution function $\phi$ and the cumulative distribution function $\Phi$ of the normal distribution $N(\mu, \sigma)$ in Equation (\ref{eq:sn}) are given by
\begin{align}
	\phi\(\frac{x-\mu}{\sigma}\) &= \frac{1}{\sqrt{2\pi}} \exp\left[-\frac{1}{2} \(\frac{x-\mu}{\sigma}\)^2\right] \\
	\Phi\(\frac{x-\mu}{\sigma}\) &= \frac{1}{2} \left[1+ \erf\(\frac{x-\mu}{\sigma\sqrt{2}}\)\right],
\end{align}
where the error function is 
\begin{equation}
	\erf(x) = \frac{2}{\sqrt{\pi}} \int_0^x e^{-t^2} dt.
\end{equation}
Then the skewed normal distribution function in Equation (\ref{eq:sn}) can be written as
\begin{equation}
	\text{SN}(M \, | \, \mu, \sigma, \alpha) = \frac{1}{\sigma\sqrt{2\pi}} \ \exp\left[-\frac{1}{2} \(\frac{M-\mu}{\sigma}\)^2\right] \nonumber \\ \times \left\{1+ \erf\left[\frac{(M-\mu)\alpha}{\sigma\sqrt{2}}\right]\right\}
\end{equation}

The asymmetric normal distribution function with $c, d > 0$ given by Equation (\ref{eq:an}) is:
\begin{eqnarray} 
	\AN(x~|~c,d) &=& \frac{2}{d\(c+1/c\)} ~ \left[ \phi\(\frac{x}{cd}\) \Theta(x) + \phi\(\frac{cx}{d}\) \Theta(-x) \right] \\
	&=&
	\begin{dcases}
		\frac{2}{d\(c+1/c\)} \frac{1}{\sqrt{2\pi}} \exp \left[{-\frac{1}{2}\(\frac{x}{cd}\)^2}\right] &\text{ if } x \ge 0 \\
		\frac{2}{d\(c+1/c\)} \frac{1}{\sqrt{2\pi}} \exp \left[{-\frac{1}{2}\(\frac{cx}{d}\)^2}\right] &\text{ otherwise },
	\end{dcases}
\end{eqnarray}
where $\phi$ is the density function of the standard normal distribution $N(0,1)$ and $\Theta$ is the Heaviside step function.

\section{Calculation of c, d}
\label{app:b}
Here we show how coefficients $c_{i,j}$, $d_{i,j}$ for the $j$th star in the $i$th population are calculated. Here we suppress the notations $i, j$ for simplicity. Let $l$ and $u$ be the 68\% lower and upper limits of the NS mass (from data), respectively. The condition
\begin{equation}
	\AN(-l \,|\, c, d) = \AN(u \,|\, c, d),
\end{equation}
implies that 
\begin{eqnarray}
	\exp \left[ {-\frac{1}{2} \( -\frac{cl}{d} \)^2} \right] = \exp \left[ {-\frac{1}{2}\(\frac{u}{cd}\)^2} \right],
\end{eqnarray}
which results into
\begin{equation} \label{eq:c}
	c = \sqrt{\frac{u}{l}}.
\end{equation}
Given $c$ by Equation (\ref{eq:c}), now we want to solve for $d$ such that
\begin{equation} \label{eq:intan}
	\int_{-l}^u \AN(x \,|\, c,d) \; dx = 0.68,
\end{equation}
where $c\equiv \sqrt{u/l}$. Note that the first-order derivative of the error function (without and with a scale factor $a$) are, respectively, 
\begin{equation}
	\frac{d}{dx} \erf(x) = \frac{2}{\sqrt{\pi}} e^{-x^2}; \quad
	\frac{d}{dx} \erf(ax) = \frac{2a}{\sqrt{\pi}} e^{-(ax)^2}. 
\end{equation} 
The last result can be multiplied by a factor to obtain:
\begin{equation} \label{eq:an3}
	\frac{1}{d\big(c+1/c\big)} \frac{1}{a\sqrt{2}} \; \frac{d}{dx} \erf(ax) = \AN(x \,|\, c,d). 
\end{equation}
Here, the scale factor $a$ is defined as:
\begin{align} \label{eq:alg}
	a = 
	\begin{dcases}
		\frac{c}{\sqrt{2}d} &= a_< \quad \text{ if } x<0, \\
		\frac{1}{\sqrt{2}cd} &= a_> \quad \text{ if } x \ge 0.
	\end{dcases}
\end{align}
Substituting Equation (\ref{eq:an3}) into (\ref{eq:intan}):
\begin{eqnarray}
	\frac{1}{\sqrt{2}a_<d(c+1/c)} \int_{-l}^0 dx \; \frac{d}{dx} \erf(a_<x) 
	+ \frac{1}{\sqrt{2}a_>d(c+1/c)} \int_0^u dx \; \frac{d}{dx} \erf(a_>x) = 0.68,
\end{eqnarray}
we obtain
\begin{equation}
	c^2\erf(a_>u) - \erf(-a_<l) = 0.68(c^2+1).
\end{equation}
Plugging in $a$ from Equation (\ref{eq:alg}), we finally have
\begin{equation} \label{eq:eqd}
	c^2\erf\(\frac{u}{\sqrt{2}cd}\) - \erf\(-\frac{cl}{\sqrt{2}d}\) - 0.68\(c^2+1\) = 0
\end{equation}
Given $c$, Equation (\ref{eq:eqd}) can be used to solve for $d$.

%% For this sample we use BibTeX plus aasjournalv7.bst to generate the



\end{document}